\documentclass[journal]{IEEEtran}
\IEEEoverridecommandlockouts

\usepackage{cite}
\usepackage{amsmath,amssymb,amsfonts}
\usepackage{graphicx}
\usepackage{textcomp}
\usepackage{xcolor}
\usepackage[linesnumbered,ruled,vlined]{algorithm2e}
\usepackage{xfrac}
\usepackage{multirow}
\usepackage{subcaption}
\usepackage{scalerel}
 \def\BibTeX{{\rm B\kern-.05em{\sc i\kern-.025em b}\kern-.08em
     T\kern-.1667em\lower.7ex\hbox{E}\kern-.125emX}}
\begin{document}
\title{Energy-Efficient Trajectory Design of a Multi-IRS Assisted Portable Access Point \\
\author{Nithin Babu,~\IEEEmembership{Student Member,~ IEEE,}\,
       Marco Virgili,~\IEEEmembership{Student Member, ~IEEE,}\,\\
       Mohammad Al-jarrah,~\IEEEmembership{Student Member,~ IEEE,}\,
       Xiaoye Jing,~\IEEEmembership{Student Member, ~IEEE,}\,\\
       Emad Alsusa, \,\IEEEmembership{~Senior member, ~IEEE,}
       Petar Popovski,\,\IEEEmembership{~Fellow, ~IEEE},\\
       Andrew Forsyth,\,\IEEEmembership{~Senior member, ~IEEE},\, Christos Masouros,\,\IEEEmembership{~Senior member, ~IEEE}, and \\ Constantinos B. Papadias,~\IEEEmembership{~Fellow,~IEEE.} 
\thanks{ Copyright (c) 2015 IEEE. Personal use of this material is permitted. However, permission to use this material for any other purposes must be obtained from the IEEE by sending a request to pubs-permissions@ieee.org. This work is supported by the project PAINLESS which has received funding from the European Union’s Horizon 2020 research and innovation programme under grant agreement No 812991.}
\thanks{N. Babu and C. B. Papadias are with Research, Technology and Innovation Network (RTIN), Alba, 
The American College of Greece, Greece (e-mail: \{nbabu, cpapadias\}@acg.edu).}
\thanks{N. Babu, C. B. Papadias, and P. Popovski are with Department of Electronic Systems, Aalborg University, Denmark (e-mail: \{niba, copa,petarp\}@es.aau.dk).}
\thanks{M. Virgili is with Lyra Electronics (e-mail: marco@lyraelectronics.com).}
\thanks{M. Virgili, M. Al-jarrah, A. Forsyth, and E. Alsusa are with The University of Manchester (e-mail: marco.virgili@postgrad.manchester.ac.uk, \{mohammad.al-jarrah, e.alsusa, andrew.forsyth\}@manchester.ac.uk).}
\thanks{X. Jing and C. Masouros are with University College London (e-mail: \{x.jing, c. masouros\}@ucl.ac.uk).}}
}
\markboth{This version has been accepted as a paper in the IEEE Transactions on Vehicular Technology}%
{Shell \MakeLowercase{\textit{et al.}}: Bare Demo of IEEEtran.cls for IEEE Journals}
\maketitle
 \begin{abstract}
 In this work, we propose a framework for energy efficient trajectory design of an unmanned aerial vehicle (UAV)-based portable access point (PAP) deployed to serve a set of ground nodes (GNs). In addition to the PAP and GNs, the system consists of a set of intelligent reflecting surfaces (IRSs) mounted on man-made structures to increase the number of bits transmitted per Joule of energy consumed measured as the global energy efficiency (GEE). The GEE trajectory for the PAP is designed by considering the UAV propulsion energy consumption and the Peukert effect of the PAP battery, which represents an accurate battery discharge profile as a non-linear function of the UAV power consumption profile. The GEE trajectory design problem is solved in two phases: \color{black}in the first, a path for the PAP and feasible positions for the IRS modules are found using a multi-tier circle packing method, and the required IRS phase shift values are calculated using an alternate optimization method that considers the interdependence between the amplitude and phase responses of an IRS element; in the second phase, the PAP flying velocity and user scheduling are calculated using a novel multi-lap trajectory design algorithm. Numerical evaluations show that: neglecting the Peukert effect overestimates the available flight time of the PAP; after a certain threshold, increasing the battery size reduces the available flight time of the PAP; the presence of IRS modules improves the GEE of the system compared to other baseline scenarios; the multi-lap trajectory saves more energy compared to a single-lap trajectory developed using a combination of sequential convex programming and Dinkelbach algorithm.    
 \end{abstract}
\section{Introduction}
An unmanned aerial vehicle (UAV) carrying a radio access node, hereafter referred to as `portable access point' (PAP), has been envisioned as a viable solution to save energy or improve user fairness in an Internet-of-Things (IoT) or federated learning application \cite{survey_mozaffari} \cite{igornbabu}. Moreover, significant progress has been made in the standardization efforts of the Third Generation Partnership Project (3GPP) to define the specifications to utilize aerial platforms for 5G and beyond \cite{3gpp}. The portable feature of a PAP can improve the communication channels of the users, \color{black}but it is limited by its finite on-board available energy. \color{black}Hence, the deployment of a PAP to serve a set of users should maximize the number of bits transmitted per Joule of energy consumed, defined as the global energy efficiency (GEE) of the system \cite{babu1}. \color{black} One of the goals of this study is the maximization of the GEE, which is achieved with a combination of two sub-goals: decreasing the energy consumption of the PAP while increasing the data rate to users. The first goal exploits the fact that a UAV consumes less energy when flying horizontally at an optimal velocity than when it hovers \cite{rui1}, \cite{babu3}. The data rate to a user is increased by improving the probability of having a line-of-sight (LoS) channel with the user by flying the PAP closer to it \cite{plos}. However, this strategy also increases the energy consumption of the UAV. Moreover, the flight time is reduced further by the Peukert effect \cite{peukert1}, \cite{peukert2}, according to which the voltage drop of a PAP's on-board battery is a non-linear function of the power output.
\begin{figure}{}
\includegraphics[width=0.9\linewidth]{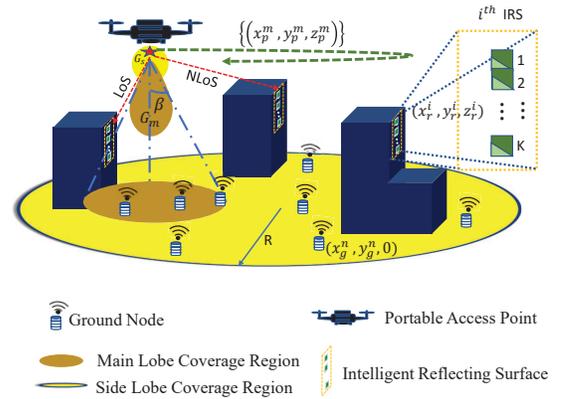}
\caption{PAP deployment scenario.}\label{system_model}
\end{figure}

Another way of enhancing the received signal power is represented by intelligent reflecting surfaces (IRSs), a key technology that provides additional paths between the transmitter and the receiver \cite{IRS-10}. An IRS is a two-dimensional surface of a finite number of elements made of a meta-material whose properties can be reconfigured using a controller\cite{emil1}. These elements are sub-wavelength-sized and can apply phase shifts on the incident waves before re-radiating them to the receiver. The direction at which the re-radiated waves add constructively can be controlled by applying suitable bias voltages to the IRS elements using the controller. However, a recent study shows that guaranteeing a constructive interference of signals at the receiver might not always improve the received signal-to-noise ratio (SNR) due to the interdependence of amplitude and phase values of a re-radiated wave from an IRS element \cite{rui_amp}. Consequently, the careful addition of IRSs to a PAP system further enhances the received signal at the user end, thereby improving the GEE of PAP communication systems. The main challenge in adding IRSs to a PAP system is finding optimal locations for the IRSs. For instance, a random IRS placement policy would place an IRS in the non-line-of-sight regime of a ground user, thereby limiting its contribution to the GEE improvement. Here we propose a method to find locations for IRSs that guarantee LoS PAP-IRS links, with each user having at least one LoS link to an IRS.\color{black} 
\subsection{Related works}
\color{black}The placement optimization of a UAV-based system has been extensively studied in the literature \cite{igornbabu}-\cite{survey2}.  In \cite{igornbabu}, the authors consider a UAV system deployed to assist slow-learning nodes in a federated learning application. The trajectory optimization problem, formulated to minimize learning time discrepancy among the nodes, is solved using a deep reinforcement learning technique and the sequential convex programming (SCP) technique. The work in \cite{babu1} finds the optimal hovering altitude of a single-UAV system that maximizes the GEE of the system, whereas \cite{babu3} considers a fly-hover-communicate protocol to serve a set of ground IoT nodes. The authors of \cite{rui1} design a trajectory for a rotary-wing UAV that minimizes the UAV propulsion energy consumption. In \cite{plos}, the authors propose a general probabilistic LoS-non-LoS (NLoS) air to ground channel model and determine the optimal altitude that maximizes the coverage region. The authors of \cite{joint} propose a graph-based algorithm to improve the throughput by jointly optimizing the user association, UAV altitude, and transmission direction. In \cite{survey1} and the references therein, the authors summarize the works that have considered UAV(s) placement problems from an energy efficiency perspective, whereas \cite{survey2} outlines the works that position UAV(s) to maximize communication-related parameters such as coverage area and throughput.  

Significant efforts have been dedicated by researchers to assess the performance of IRS panels in many wireless applications, such as multi-hop integrated access and backhauling (IAB) \cite{IRS-4}, localization, physical layer security, and simultaneous wireless and information power transfer (SWIPT) \cite{IRS-3}. The works in \cite{IRS-5}-\cite{IRS-WB-1} consider IRS for aiding the communication in a UAV-based system. In \cite{IRS-5}, the effect of phase error compensation on the achievable error rate and outage probability is investigated for UAV-IRS systems, and its impact on the achievable capacity is analyzed in \cite{IRS-6}. However, the trajectory design of UAVs is not taken into account in these works. In \cite{IRS-7-a} and \cite{IRS-7}, IRS panels are deployed on high buildings to assist the downlink transmission from a flying base station to a single ground user. The authors consider beamforming at IRS and UAV trajectory design to maximize the average achievable rate and the power received by the user, respectively. The sum-rate maximization problem is addressed in \cite{IRS-8}, where the joint design of IRS beamforming, IRS scheduling, and UAV trajectory have been considered. In \cite{IRS-9}, the weighted sum bit error rate (BER) achieved by multiple IRSs is minimized by jointly optimizing the IRS phase shift matrix, the UAV trajectory, and the scheduling of the IRSs. The UAV trajectory design for a UAV-IRS system operating in terahertz (THz) band is considered in \cite{IRS-WB-1}, where a single IRS panel with ideal phase compensation is assumed. The authors of \cite{petar} propose a continuous-time system model for multi-path channels and discuss the optimal IRS configuration with respect to the received power, Doppler spread, and delay spread. Recent studies have also considered the possibility of using backscattering to aid the communication between the nodes\cite{rev1}\cite{rev2}. The reconfigurability of IRSs makes them suitable for a UAV-based system.  \color{black}
\subsection{Main contributions and paper organization}
The work in \cite{igornbabu}, \cite{rui1}-\cite{survey2} and the references therein consider either the maximization of communication-related parameters (sum rate, coverage area), or the minimization of the energy consumed in a UAV-based system. \cite{babu1} and \cite{babu3} propose UAV placement policies to maximize the GEE that are suitable for hovering and fly-hover-communicate scenarios, respectively. Here we allow the PAP to serve the users while it is flying. Additionally, the works in \cite{IRS-4}-\cite{petar}
consider scenarios with either a single user with a single IRS panel \cite{IRS-7-a}, a single IRS with multiple users \cite{IRS-8}, or multiple IRSs with a single user \cite{IRS-7,IRS-9} to maximize the sum rate or minimize the BER. \color{black}Moreover, \cite{IRS-8} and \cite{IRS-WB-1} consider the trajectory design of UAVs assisted by a single IRS operating in a wide-band setting. However, a single IRS with ideal reflection and perfect phase compensation is assumed. Unlike the existing literature, we consider a generalized system model with a multi-user multi-IRS scenario to maximize the number of bits transmitted per Joule of energy consumed. Moreover, practical limitations for IRS design are considered, such as the phase-amplitude relation and discrete phase compensation. It is worth highlighting that none of the above-mentioned work considers the \color{black} non-linearity introduced by the Peukert effect on the battery discharge profile of UAVs. Additionally, the relationship between the number of cells of the on-board battery and the maximum flight time of a UAV has not been investigated in the literature. Here we consider the 3GPP-proposed air-to-ground channel model to estimate the UAV-IRS and IRS-user path gains that are later used for IRS positioning. The UAV-user path gain is estimated using the widely accepted LoS-NLoS path loss model proposed in \cite{plos}. Our main contributions are summarized as follows:
\begin{itemize}
    \item A discharge characteristic for Li-ion batteries is obtained by applying a non-linear regression analysis to the discrete data provided in the battery data-sheet. This way, the initial and final voltage and the maximum capacity can be evaluated for all current values with great accuracy. \color{black} Such an empirical approach differs from analytical models like those in \cite{peukert1} and \cite{peukert2}. 
    \item An algorithm to estimate the available flight time of a PAP considering the Peukert effect of the PAP battery, which is usually neglected in studies involving UAVs. \color{black}For a given flying velocity, the estimation is done by representing the required power as a function of the battery terminal voltage and current using the developed discharge characteristic. Also, we investigate the trade-offs of adding more battery cells to the PAP battery by considering its positive (larger initial capacity) and negative effects (heavier PAP).
    \item \color{black}IRS positioning guidelines are introduced considering the 3GPP air-to-ground channel model with the proposed multi-tier circle packing algorithm. Moreover, we optimize the additional phase shift introduced by the elements of an IRS, considering the interdependence of its amplitude and phase responses. 
    \item \color{black}The estimated available flight time and the determined IRS positions are then utilized in developing an energy-efficient path planning (E2P2) algorithm for a PAP deployed to serve a multi-user multi-IRS system. The algorithm maximizes the global energy efficiency of the system by considering the UAV propulsion energy consumption and the 3GPP air-to-ground channel.
    
\end{itemize}
 \color{black}
The remainder of this paper is organised as follows. In Section~\ref{sysmodel}, we describe the scenario under consideration, and explain the propagation environment and the PAP power consumption model. Building on this, Section~\ref{optimaldesign} starts by explaining the Peukert effect, then proposes an algorithm to estimate the available flight time of a PAP as a function of its velocity. Then, in Section~\ref{prblm}, the GEE maximization problem is formulated and solved with a two-phase approach described in Section~\ref{Phase 1} and Section~\ref{phase2}. The main findings of the numerical evaluation are reported in Section~\ref{result}.
\section{System Model and Definitions}\label{sysmodel}
In this work, we consider an unmanned aerial system (UAS) in which a PAP is deployed to deliver $Q$ bits of data to a set of $N$ outdoor ground nodes (GNs) located at $\mathbf{g}_n=[x^{n}_{\mathrm{g}},y^{n}_{\mathrm{g}},0]$, $n\in \mathcal{N}=\{1,2,3,..,N\}$.  
In addition to the PAP and the GNs, a set of $I$ IRS modules are deployed. The presence of IRS modules (IRSs) could aid the communication between the PAP and the GNs by providing additional paths for the signal from the PAP to reach the GNs. This improves the GEE of the system by increasing the received signal power at the GNs, thereby reducing the total mission time and energy consumption of the PAP. The PAP is assumed to fly horizontally at an altitude $h_\mathrm{p}$. Additionally, we assume the PAP to be equipped with a directional antenna, the gain of which, in the direction $(\alpha, \epsilon)$, is given by, 
\begin{IEEEeqnarray}{rCl}\label{ga}
G_{\mathrm{a}}& = & \left\{\begin{matrix}
G_{\text{m}}&-\beta\leq\alpha\leq\beta, -\beta\leq\epsilon\leq\beta,\\
G_{\text{s}} &\text{otherwise},
\end{matrix}\right.
\label{antenna Gain}
\end{IEEEeqnarray}
where $G_{\text{m}}={2.2846}/{\beta^{2}}$, and $G_{\text{s}}$ are the main and side lobe gains of the PAP antenna, respectively \cite{joint}; the half-power beamwidth of the antenna in the elevation and the azimuth plane is $2\beta$. 
The GNs are considered to be equipped with omni-directional antennas. 

\color{black}The trajectory optimization is assumed to happen offline at the ground station prior to the PAP deployment. This requires the ground station to be aware of the positions of the GNs, which can be done, for instance, using the new radio positioning protocol (NRPPa). We consider a scenario where the ground nodes are static, which relates to practical sensor-centric IoT scenarios. Hence, the offline computed path could be used throughout the mission. Based on the GN scheduling, the bias voltages of the elements of IRSs are varied through control channels existing between the PAP and the IRSs. Furthermore, we assume both the PAP and GNs to be aware of the channel state information, and the backhaul link for the PAP is achievable with the new integration of low earth orbit (LEO) satellites \cite{satbackhaul}. Hence, not considered in the analysis. \color{black}
\subsection{PAP trajectory model}
For tractability, the total flying path of the PAP is divided into $M$ segments, represented using $M+1$ way points, whose locations are denoted as $\mathbf{p}_m=[x^{m}_{\mathrm{p}},y^{m}_{\mathrm{p}},z^{m}_{\mathrm{p}}]$, $m\in \mathcal{M}=\{1,2,3,..,M+1\}$: $z_\mathrm{p}^m=h_\mathrm{p}$ $\forall m$. The length of each segment is constrained to be small enough as to leave the channel between the PAP and ground modules (IRS and GNs) unchanged, while the PAP is in a given path segment \cite{rui1}:
\begin{IEEEeqnarray}{rl}
\begin{Vmatrix}\mathbf{p}_{m+1}-\mathbf{p}_m\end{Vmatrix}\leq\text{min}\left\lbrace \Delta, T_{m} v_{\text{max}}\right\rbrace  \,\,\forall m \in \mathcal{M}^{'}\label{path_descritization},
\end{IEEEeqnarray}
where $\mathcal{M}^{'}=\mathcal{M}-\{M+1\}$; the segment length $\Delta$ is appropriately chosen so that, within each line segment, the PAP can be assumed to fly with a constant velocity $v_m$, and the distances between the PAP and each GN and IRS modules are approximately unchanged: $\Delta<<h_\mathrm{p}$; let $T_m$ be the time which the PAP spends in the $m^{\text{th}}$ path segment and $v_{\text{max}}$ be the maximum horizontal flying velocity of the PAP. In any given segment, the PAP follows a time-division multiple access (TDMA) scheme to serve the GNs: let $T_{mn}$ be the time allocated to the $n^{\text{th}}$ GN while the PAP is in the $m^{\text{th}}$ path segment such that,
\begin{IEEEeqnarray}{rl}
\Sigma_{n=1}^{N}T_{mn} \leq T_{m} \quad \forall m\in \mathcal{M}^{'}. \label{tdma}
\end{IEEEeqnarray}
\subsection{IRS model}
\begin{figure}{}
\includegraphics[width=0.9\linewidth]{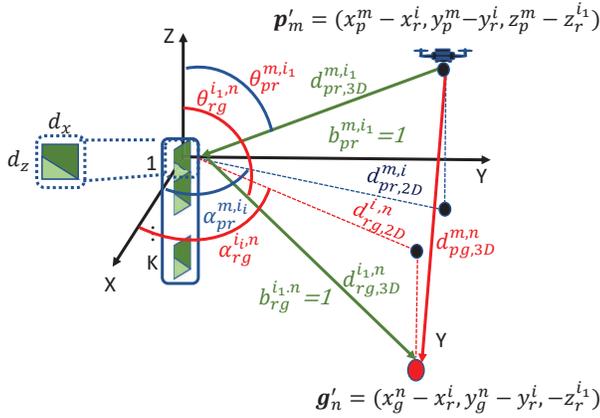}
\caption{Propagation environment considering the $m^{\text{th}}$ path segment, $i^{\text{th}}$ IRS and $n^{\text{th}}$ GN. }\label{prop_model}
\end{figure}

Each IRS is considered to be a uniform linear array (ULA) of $K$ reflecting elements with dimensions $d_\mathrm{x}$ and $d_\mathrm{z}$, placed along the positive $Z$ axis as shown in Fig.~\ref{prop_model}. 
The first element of each IRS is considered as the reference element and its geometric center is having the coordinates $\mathbf{r}_i=[x^{i}_{\mathrm{r}},y^{i}_{\mathrm{r}},z^{i}_{\mathrm{r}}]$, $i\in \mathcal{I}=\{1,2,3,..,I\}$. Then, the coordinates of the $k^{\text{th}}$ reflecting element of the $i^{\text{th}}$ IRS are $\mathbf{r}_i^{k}=(x_\mathrm{r}^i,y_\mathrm{r}^i,(z_\mathrm{r}^i-(k-1)d_\mathrm{z}))$ $\forall k\in [1, K], \forall i \in \mathcal{I}$. Additionally, while considering the channel between the PAP and the $n^{\text{th}}$ GN, if either of them is located behind the $i^{\text{th}}$ IRS, the IRS is not considered for transmission to the GN. Let $\{b_{\mathrm{pr}}^{m,i}\}$ and $\{b_{\mathrm{rg}}^{i,n}\}$ be the respective binary variables whose value is 1 if the PAP at the $m^{\text{th}}$ segment and the $n^{\text{th}}$ GN are in front of the $i^{\text{th}}$ IRS, respectively:
 \begin{IEEEeqnarray}{rCl}\label{bpr}
b_{\mathrm{pr}}^{m,i}& = & \left\{\begin{matrix}
1 &\text{if}\,\,\alpha^{m,i}_{\mathrm{pr}}=\text{arctan}\left[\dfrac{(y^{m}_{\mathrm{p}}-y^{i}_{\mathrm{r}})}{(x^{m}_{\mathrm{p}}-x^{i}_{\mathrm{r}})}\right]\in [0,-1^{\psi}\pi],\\
0&\text{otherwise},
\end{matrix}\right.
\end{IEEEeqnarray}
\begin{IEEEeqnarray}{rCl}\label{brg}
b_{\mathrm{rg}}^{i,n}& = & \left\{\begin{matrix}
1 &\text{if}\,\,\alpha^{i,n}_{\mathrm{rg}}=\text{arctan}\left[\dfrac{(y^{n}_{\mathrm{g}}-y^{i}_{\mathrm{r}})}{(x^{n}_{g}-x^{i}_{\mathrm{r}})}\right]\in [0,-1^{\psi}\pi],\\
0&\text{otherwise},
\end{matrix}\right.
\label{crg}
\end{IEEEeqnarray}
in which the ${arctan}$ function considers the correct quadrant of the argument. \color{black} Moreover, a semiconductor device, such as a PIN diode, is used to tune the impedance of a reflecting element of an IRS in real-time without changing the geometrical parameters. This can be done by controlling its biasing voltage using a controller attached to each IRS. The amplitude and phase responses of a reflecting element are mutually dependent. In \cite{rui_amp}, the authors have presented the relationship in a closed-form as,
\begin{IEEEeqnarray}{rCl}
\mu_{m,i,n}^{k}(\theta_{m,i,n}^{k}) &=& (1-\mu_{\text{min}})\cdot \left({\frac{\text{sin}(\theta_{m,i,n}^{k} - \varrho) + 1}{2}}\right)^{\zeta} + \mu_{\text{min}},\nonumber\\\label{amp-phase}
\end{IEEEeqnarray}
where $\mu_{\text{min}}\geq 0$, $\varrho\geq0$, and $\zeta \geq 0$ are the constants related to the circuit implementation of the reflecting element. Let $ \mathbf{\Theta}_{\mathrm{r}}^{m,i,n} =\text{diag}\left( \mu_{m,i,n}^{1}(\theta_{m,i,n}^{1})e^{j\theta_{m,i,n}^{1}},.., \mu_{m,i,n}^{K}(\theta_{m,i,n}^{K})e^{j\theta_{m,i,n}^{K}}\right),  i\in\mathcal{I}$, $\theta_{m,i,n}^{k} \in \left[0,\pi\right)$, be the amplitude-phase shift matrix of the $i^{\text{th}}$ IRS when the PAP serves the $n^{\text{th}}$ GN from the $m^{\text{th}}$ segment.\color{black} 
\subsection{Propagation Environment}\label{prop}
\color{black}We consider the system to be deployed in an urban environment where the air-to-ground link can be either LoS or NLoS, depending on the blockage profile of the environment and the relative position of the receiver module (IRS/GN) to the transmitter module (PAP/IRS) \cite{plos}, \cite{3GPPlte}\color{black}. Consequently, the mean path loss value of an air-to-ground link has the form, 
\begin{IEEEeqnarray}{rCl}
\overline{L}^{c_2,c_3}_{c_1}=\underbrace{P_{c1,{1}}^{c2,c3}\times L_{c_1,{1}}^{c_2,c_3}}_{\text{LoS Pathloss}}+\underbrace{(1-P_{c_1,{1}}^{c_2,c_3})\times L_{c_1,{2}}^{c_2,c_3},}_{\text{NLoS Pathloss}}
\end{IEEEeqnarray}
in which $c_1\in \{\mathrm{pr}, \mathrm{rg} \}$, $c_2\in \{m,i_k\}$, $c_3\in \{i_k, n\}$, where $i_{k}$ represents the $k^{\text{th}}$ element of the $i^{\text{th}}$ IRS module.  
\subsubsection{PAP-IRS channel}
From the 3GPP report \cite{3GPPlte}, the LoS and NLoS path loss values between the $m^{\text{th}}$ path segment and the $k^{\text{th}}$ element of the $i^{\text{th}}$ IRS can be expressed as,
\begin{IEEEeqnarray}{rCl}
L_{\mathrm{pr},\text{1}}^{m,i_k}& = & 
30.9+(22.25-0.5\text{log}h_\mathrm{p})\text{log}d_{\mathrm{pr},\text{3D}}^{m,i_k}+F\label{pathlosspr1}, \\
L_{\mathrm{pr},\text{2}}^{m,i_k}&=&\text{max}\{L_{\mathrm{pr},1}^{m,i_k},32.4+(43.2-7.6\text{log}h_\mathrm{p})\text{log}d_{\mathrm{pr},\text{3D}}^{m,i_k}
+F\}, \nonumber\label{pathlosspr2}\\
\end{IEEEeqnarray}
where $F=20\text{log}f$, with $f$ being the carrier frequency; $d_{\mathrm{pr},\text{3D}}^{m,i_k}=\|\mathbf{p}_m-\mathbf{r}_i^k\|$. Given that the IRSs are perpendicular to the ground, the azimuth angles for the waves arriving at the reflecting elements of an IRS from the PAP are equal. The corresponding LoS probability is expressed as,
\begin{IEEEeqnarray}{rCl}
P_{\mathrm{pr},\text{1}}^{m,i}& = & \left\{\begin{matrix}
1\,\,\text{if}\quad d_{\mathrm{pr},\text{2D}}^{m,i}\leq d_{1},\label{plos}\\
\dfrac{d_1}{d_{\mathrm{pr},\text{2D}}^{m,i}}+\text{exp}\left[\dfrac{-d_{\mathrm{pr},\text{2D}}^{m,i}}{p_1}\right]\left[1-\dfrac{d_1}{d_{\mathrm{pr},\text{2D}}^{m,i}}\right];\text{else,}
\end{matrix}\right.\label{plospr}
\end{IEEEeqnarray}
where,
\begin{IEEEeqnarray}{rCl}
d_{\mathrm{pr},\text{2D}}^{m,i}&=&\sqrt{(x_\mathrm{p}^m-x_\mathrm{r}^i)^2+(y_\mathrm{p}^m-y_\mathrm{r}^i)^2},\\
p_1 &=& 233.98 \text{log}_{10}(h_\mathrm{p})-0.95,\\
d_1 &=& \text{max}\left( 294.05\cdot\text{log}_{10}(h_\mathrm{p})-432.94,18\right).\label{d1pr}
\end{IEEEeqnarray}
Also, \eqref{pathlosspr1}-\eqref{d1pr} are valid when $22.5< h_\mathrm{p} \leq 100$m and when the IRSs are located at a height of 10m (for an urban scenario).
It is worth pointing out that most of the works in the literature, when considering the relative phase of the incident wave on the ULA elements, assume that the reference element has a phase of $0^{o}$. However, since multiple IRSs are considered in this work, each with different location and reference points, the actual phase should be considered. Hence, the channel gain vector  between the $i^{\text{th}}$ IRS and the PAP while the PAP is in the $m^{\text{th}}$ path segment is represented as,
\begin{multline}
\mathbf{h}_{\mathrm{pr}}^{m,i} = \left(b_{\mathrm{pr}}^{m,i}\sqrt{G_{\mathrm{pr}}^{m,i_k}10^{-\overline{L}^{m,i_k}_{\mathrm{pr}}/10}}  e^{-j\frac{2\pi}{\lambda}d_{\mathrm{pr},\text{3D}}^{m,i_k}}\right)_{k={1,...,K}}
\end{multline}
where $G_{\mathrm{pr}}^{m,i_k}=G_{\text{m}}$ if $\text{arctan}[|((z_\mathrm{r}^i-(k-1)d_\mathrm{z})-z_{\mathrm{p}}^{m})|/d_{\mathrm{pr},\text{2D}}^{m,i}]\leq \beta$; else $G_{\text{s}}$.
\subsubsection{IRS-GN channel}
Similarly to the previous sub-section, from \cite{3GPPlte}, the LoS and NLoS path loss values between the elements of the $i^{\text{th}}$ IRS and the $n^{\text{th}}$ GN are estimated using,
\begin{IEEEeqnarray}{rCl}
L_{\mathrm{rg},1}^{i_k,n}& = & \left\{\begin{matrix}
L_{1} \quad \quad \text{if} \quad 10\text{m}\leq d^{i_k,n}_{\mathrm{rg},2D}\leq d_{BP},\\
L_{2} \quad \quad \text{if} \quad d_{BP} \leq d^{i_k,n}_{2D}\leq 5\text{km},
\end{matrix}\right.
\end{IEEEeqnarray}
\begin{IEEEeqnarray}{rCl}
L_{\mathrm{rg},2}^{i_k,n}& = & \text{max}\left(L_{\mathrm{rg},1}^{i_k,n},{L_{\mathrm{rg},2}^{'i_k,n}}\right) \,\, \text{for}  10\text{m}\leq d^{i_k,n}_{\mathrm{rg},2D}\leq 5\text{km}\nonumber
\end{IEEEeqnarray}
where $L_{1}=32.4+21\text{log}(d^{i_k,n}_{\mathrm{rg},3D})+F$ and $L_{2}=32.4+40\text{log}(d^{i_k,n}_{\mathrm{rg},3D})+F-9.5\text{log}\left[(d_{BP})^2+72.25\right]$ with $d_{BP}=18f /c$, $c=3\times 10^8$m/s; $L_{\mathrm{rg},2}^{'i_k,n}=35.3\text{log}_{10}(d^{i_k,n}_{\mathrm{rg},3D})+22.4+21.3\text{log}_{10}(f)$. The LoS probability is determined using,
\begin{IEEEeqnarray}{rCl}
P_{\mathrm{rg},1}^{i,n}& = & \left\{\begin{matrix}
1\,\,\text{if}\quad d_{\mathrm{rg},\text{2D}}^{i,n}\leq 18 \text{m},\nonumber\\
\dfrac{18}{d_{\mathrm{rg},\text{ 2D}}^{i,n}}+\text{exp}\left[\dfrac{-d_{\mathrm{rg},\text{2D}}^{i,n}}{36}\right]\left[1-\dfrac{18}{d_{\mathrm{rg},\text{2D}}^{i,n}}\right];\text{else.}
\end{matrix}\right.\nonumber\\\label{plosrg}
\end{IEEEeqnarray}
 The channel gain between the $i^{\text{th}}$ IRS and the $n^{\text{th}}$ GN is given by:
 \begin{multline}
\mathbf{h}_{\mathrm{rg}}^{i,n} = \left(b_{\mathrm{rg}}^{i,n}\sqrt{10^{-\overline{L}^{i_k,n}_{\mathrm{rg}}/10}}  e^{-j\frac{2\pi}{\lambda}d_{\mathrm{rg},\text{3D}}^{i_k,n}}\right)_{k={1,...,K}}.
\end{multline}
\subsubsection{PAP-GN channel}
The LoS and NLoS path loss values between the $m^{\text{th}}$ PAP path segment and the $n^{\text{th}}$ GN can be  
expressed as \cite{plos}, \cite{babu3},
 \begin{IEEEeqnarray}{rCl}\label{pathloss}
 L_{\mathrm{pg},1}^{m,n}& = & 
 20\text{log}d_{\mathrm{pg},\text{3D}}^{m,n}+F+20\text{log}\left(\dfrac{4\pi}{c}\right)+\eta_{\text{1}}\label{pathlosspg1}, \\
 L_{\mathrm{pg},2}^{m,n}& = & 
 20\text{log}d_{\mathrm{pg},\text{3D}}^{m,n}+F+20\text{log}\left(\dfrac{4\pi}{c}\right)+\eta_{\text{2},}
 \end{IEEEeqnarray}
with $d_{\mathrm{pg},\text{3D}}^{m,n}=\|\mathbf{p}_{m}-\mathbf{g}_{n}\|$. The corresponding probability of existence of a LoS link between the PAP and the GN is expressed as \cite{plos},
\begin{IEEEeqnarray}{rCl}
P_{\mathrm{pg},\text{1}}^{m,n}& = & \dfrac{1}{1+a\exp{\left[-b(\phi^{m,n}_{\mathrm{pg}}-a)\right]}}, \label{plospg}
\end{IEEEeqnarray}
with $\phi^{m,n}_{\mathrm{pg}}=\text{arctan}\left({z_{\mathrm{p}}^{m}}/{\sqrt{(x_{\mathrm{p}}^{m}-x_{\mathrm{g}}^{n})^{2} +(y_{\mathrm{p}}^{m}-y_{\mathrm{g}}^{n})^{2} }}\right)$; $a$ and $b$ are environment-dependent parameters; $\eta_{\text{1}}$ and $\eta_{\text{2}}$ are the respective additional path loss values due to long-term channel variations. 
The corresponding channel gain is expressed as,
\begin{IEEEeqnarray}{rCl}
h_{\mathrm{pg}}^{m,n}&=& \sqrt{ G_{\mathrm{pg}}^{m,n} 10^{-\overline{L}^{m,n}_{\mathrm{pg}}/10}}e^{-j\frac{2\pi}{\lambda}d_{\mathrm{pg},\text{3D}}^{m,n}}\label{directPL}.
\end{IEEEeqnarray}
where $G_{\mathrm{pg}}^{m,n}=G_{\text{m}}$ if $\phi_{\mathrm{pg}}^{m,n}\leq \beta$; else $G_{\text{s}}$.

The signal transmitted from the PAP reaches a GN through two main paths, PAP-GN link and PAP-IRS link. The received SNR value at the $n^{\text{th}}$ GN while the PAP is in the $m^{\text{th}}$ path segment is given by,
\begin{IEEEeqnarray}{rCl}
 \gamma_{\mathrm{pg}}^{m,n} &=&\dfrac{P|h^{m,n}_{\mathrm{pg}}+\sum_{i\in \mathcal{I}}{\mathbf{h}^{i,n}_{\mathrm{rg}}}^{H}\Theta_{r}^{m,i,n}\mathbf{h}^{m,i}_{\mathrm{pr}}|^{2}}{\sigma^{2}},\label{snr}
\end{IEEEeqnarray}
where $P$ is the transmitted power and $\sigma^{2}$ is the additive white Gaussian noise power. \color{black}
\color{black} Assuming the availability of channel state information (CSI) at both the transmitter and the receiver, the number of bits transmitted-per-second (bps) is given by,\color{black}
\begin{IEEEeqnarray}{rCl}
D_{\mathrm{pg}}^{m,n} &=& B_{\mathrm{c}} \text{log}_{2}\left[1+\gamma_{\mathrm{pg}}^{m,n}\right]\quad\quad \forall j\in \mathcal{N}, m\in \mathcal{M}^{'}\label{Dpg},
\end{IEEEeqnarray}
where $B_{c}$ is the available channel bandwidth for each GN.
\subsection{PAP Power Consumption Model}
\begin{table}[]
\caption{UAV Parameters \cite{babu3}.}
\centering
\begin{tabular}{lll}
\hline
Label & Definition & Value \\ \hline
\hline
$W$ & Weight of the UAV in Newton & 24.5 N \\
$N_{\text{R}}$ & Number of rotors & 4 \\
$v_m$ &  UAV's horizontal flying velocity & -\\
$v_\text{tip}$ & Tip speed of the rotor & 102 m/s \\
$A_{\text{f}}$ & Fuselage area & 0.038 $\text{m}^2$  \\
$\rho(h_{a})$ & Air density & - \\
$C_{D}$ & Drag Co-efficient & 0.9  \\
$A_{r}$ & Rotor disc area & 0.06 $\text{m}^2$ \\
$\Delta_p$ & Profile drag coefficient & 0.002 \\
$s$ & Rotor solidity & 0.05\\ \hline
\end{tabular}
\label{uavparameters}
\end{table}
Since the energy consumed by the communication unit is much lower than that consumed by the aerial vehicle, we neglect the communication energy part. The UAV parameters used in this section are summarized in Table \ref{uavparameters}. The power consumed by a rotary-wing UAV while flying horizontally with a velocity $v_m$ is determined using \cite{babu3} as,
\begin{IEEEeqnarray}{rCl}\label{phfly}
P_{\text{uav}}(v_{m})&=&\underbrace{ N_{\mathrm{R}}P_{\mathrm{b}}\left(1+\dfrac{3v^2_{m}}{v_{\text{tip}}^2}\right)}_{P_{\text{blade}}}+\underbrace{\dfrac{1}{2}C_{D}A_{\text{f}}\rho(h_\mathrm{a})v^3_{m}}_{P_{\text{fuselage}}}\nonumber\\
&+&\underbrace{ W\left(\sqrt{\dfrac{W^2}{4 N_{\mathrm{R}}^2 \rho^2 (h_\mathrm{a})A_{\mathrm{r}}^2}+\dfrac{v_{m}^4}{4}}-\dfrac{v_{m}^2}{2}\right)^{1/2}}_{P_{\text{induce}}},\label{puav}
\end{IEEEeqnarray}
where $P_{\text{b}}=\dfrac{\Delta}{8}\rho(h_\mathrm{a}) s A_{\mathrm{r}} v^3_{\text{tip}}$, $\rho(h_{\mathrm{a}})=(1-2.2558.10^{-5} h_{\mathrm{a}})^{4.2577}$; $W=W_{\text{bt}}+W_{\text{body}}$ is the total weight of the UAV, comprehensive of body and battery unit. $P_{\text{blade}}$ and $P_{\text{fuselage}}$ are the powers required to overcome the profile drag forces of the rotor blades and the fuselage of the aerial vehicle that oppose its forward movement, respectively. $P_{\text{induce}}$ represents the power required to lift the payload. The hovering power is obtained by substituting $v_{m}=0$ in \eqref{phfly}.
\subsection{\color{black}Global Energy Efficiency}\color{black}
\label{geea}
The global energy efficiency of a PAP system is defined as the total number of bits transmitted per Joule of energy consumed\cite{babu1}:
 \begin{IEEEeqnarray}{c}
\text{GEE}[\text{bits}/{\text{Joule}}]=\dfrac{\sum_{m=1}^{M}\sum_{n=1}^{N}T_{mn}D_{\mathrm{pg}}^{m,n}[\text{bits}]}{\sum_{m=1}^{M}{T}_{m}P_{\text{uav}}\left(v_{m}\right)[\text{Joule}]},
\label{gee}
\end{IEEEeqnarray}
where the numerator is the total number of data bits transmitted from the PAP to the GNs at the end of the $M^{\text{th}}$ path segment and the denominator is the total energy consumed by the PAP during its flight.\color{black}
\section{GEE PAP Trajectory Design} \label{optimaldesign}
In this section, we propose an algorithm to estimate the available flight time of a PAP, considering the Peukert effect on the UAV battery. Subsequently, we design a globally energy efficient trajectory for the PAP to deliver data to the GNs.
\subsection{Available Flight Time Estimation}\label{bat_design} 
The available flight time requires an iterative calculation because it depends on the battery discharge profile, which is a non-linear function of the power drawn by the rotors of the UAV. This non-linear behavior of the UAV battery is defined by the Peukert effect \cite{peukert1},\cite{peukert2}.
\subsubsection{Peukert Effect}
\begin{figure}{}
\centering
\includegraphics[width=0.85\columnwidth]{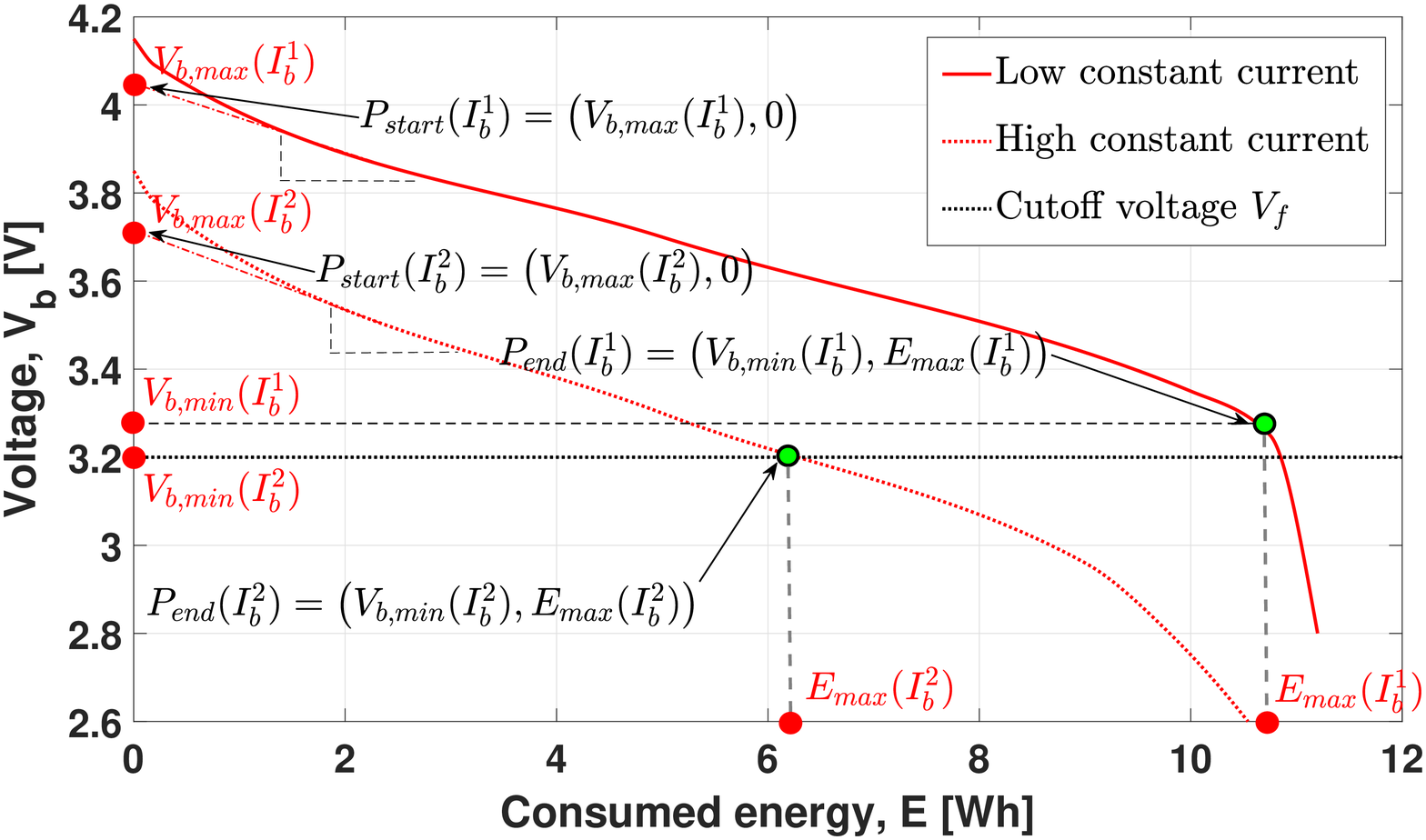}
\caption{Peukert curves for a Li-ion battery cell.}\label{fig:peukert}
\end{figure}

\begin{figure}{}
\centering
\includegraphics[width=0.85\columnwidth]{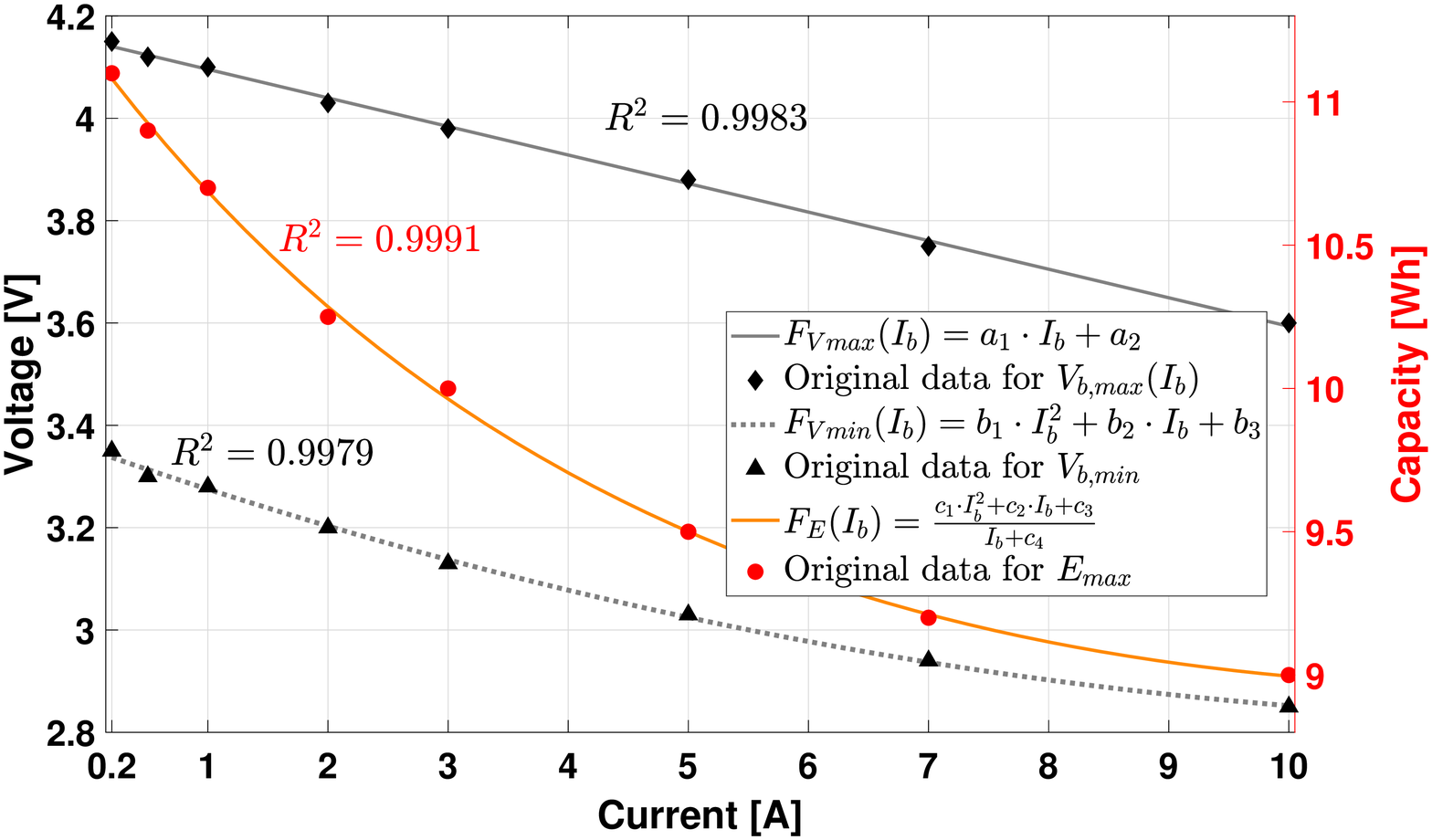}
\caption{Regression functions for $V_{b,\text{max}}$, $V_{b,\text{min}}$, and $E_{\text{max}}$.}\label{fig:regression}
\end{figure}

Fig.~\ref{fig:peukert} portrays the voltage drop of a typical Li-ion battery (commonly used in UAVs) during discharge, in various conditions. As shown in the figure, a battery is useful until the terminal voltage becomes lower than a given threshold ($V_{\text{cf}}$), or the discharge curve comes out of its linear section, whichever happens first. 
In the example reported in Fig.~\ref{fig:peukert}, the discharge at high current (dotted red curve) reaches the cutoff voltage before the end of its linear section, whereas the opposite is true at low current (continuous red curve).
The curves offer a clear explanation of the Peukert effect: as the current drawn from the battery unit increases, the available capacity (time) decreases as a non-linear function of the output current, contrary to what is often assumed in the literature \cite{rui1}. In the considered scenario, the current drawn from the battery is a function of the power consumed by the PAP, its terminal voltage, and the number of the battery cells $n_\mathrm{c}$ that form the battery unit of the PAP:
\begin{IEEEeqnarray}{rCl}
   I_\mathrm{b}^{m} &=& \dfrac{P_{\text{uav}}(v_m)}{V_\mathrm{b}^{m}\cdot n_\mathrm{c}} \quad \quad \forall m\in \mathcal{M}^{'}. \label{ibm}
\end{IEEEeqnarray}
In practice, the rates at which the battery terminal voltage drops under different conditions are determined experimentally and typically reported in the battery data sheet, like the curves of Fig.~\ref{fig:peukert}. Since data-sheets typically present a limited number of such curves, we propose to simulate this phenomenon for a continuous range of currents by adopting a hybrid approach between those developed in \cite{peukert1} and \cite{peukert2}, but based on data from data-sheets rather than analytical models. In particular, as shown in Fig.~\ref{fig:peukert}, we extrapolate discharge curves from the data provided in \cite{LGMH1}, in order to have information for several current values. The coordinates of two points are registered for each value of discharge current: the voltage and used capacity at the full charge point (taken as a continuation of the linear segment, neglecting the initial voltage drop), $P_{\text{start}}(I_\mathrm{b})=(V_{\mathrm{b},\text{max}}(I_\mathrm{b}), 0)$, and the point where the useful capacity ends, $P_{\text{end}}(I_\mathrm{b})=(V_{\mathrm{b},\text{min}}(I_\mathrm{b}), E_{\text{max}}(I_b))$. The latter corresponds to the point after which the curve's slope cannot be considered constant anymore, as shown in Fig.~\ref{fig:peukert}. Between $P_{\text{start}}$ and $P_{\text{end}}$, the slope of change of battery terminal voltage is calculated as, 
\begin{IEEEeqnarray}{rCl}\label{slopeIb}
    k(I_\mathrm{b})=\frac{V_{\mathrm{b},\text{max}}(I_\mathrm{b})-V_{\mathrm{b},\text{min}}(I_\mathrm{b})}{E_{\text{max}}(I_b)}.
\end{IEEEeqnarray}
A regression analysis is then carried out using the `fit' function in MATLAB  to generate three functions of the output current: $F_{\text{Vmax}}$, $F_{\text{Vmin}}$, and $F_{\mathrm{E}}$. These are used to determine the initial and final voltage points, and the maximum capacity for any current value. As reported in Fig.~\ref{fig:regression}, the maximum voltage is a linear decreasing function $F_{\text{Vmax}}$ of current, while $F_{\text{Vmin}}$ is quadratic and the capacity function $F_\mathrm{E}$ has a rational formulation. All three functions represent a great fit for the respective original data, with values of $R^2$ (an indicator of `goodness-of-fit' ranging from 0 to 1) next to 1.

If the PAP consumes $P_{\text{uav}}$ Watts at all times while active, the available flight time can be estimated using Algorithm \ref{algorithm2}. The required power is assumed to be drawn equally from the $n_\mathrm{c}$ battery cells. The algorithm starts by initializing the parameters of a cell as $V_\mathrm{b}^1=V_\mathrm{b}^0$, the rated terminal voltage of a battery cell and current $I_\mathrm{b}^1$ determined using \eqref{ibm}. The total flight time is divided into chunks of small intervals $\Delta t$, so that the battery terminal voltage can be assumed constant during this interval. The first operation is the calculation of the slope $k$ of the discharge curve, using \eqref{slope}:
\begin{IEEEeqnarray}{rCl}\label{slope}
    k(I_\mathrm{b}^j)=\frac{F_{\text{Vmax}}(I_\mathrm{b}^j)-F_{\text{Vmin}}(I_\mathrm{b}^j)}{F_{\mathrm{E}}(I_\mathrm{b}^j)}.
\end{IEEEeqnarray}
The parameter $k(I_\mathrm{b}^j)$ is then used to calculate the voltage at instant $(j+1)$, which, after verifying it is not lower than the cutoff voltage, allows to calculate the current at the $(j+1)^{\text{th}}$ time step. It should be noted that even though the power requirement from the PAP remains the same, the current drawn from each battery cell increases after each step due the drop in terminal voltage. This allows us to calculate the energy consumed in the next time step $E_b^{j+1}$, the total energy consumed in the mission $E_{\text{tot}}$ up to the current slot, and to re-evaluate the maximum available energy $E_{\text{max}}(I_b^{j+1})$. If the total energy consumed is greater than the maximum available energy, the loop stops and the maximum flight time $T(P_{\text{uav}})$ is obtained by multiplying the length of a time step, $\Delta t$, with the number of iterations, $j$.

\begin{figure}{}
\centering
\includegraphics[width=0.85\columnwidth]{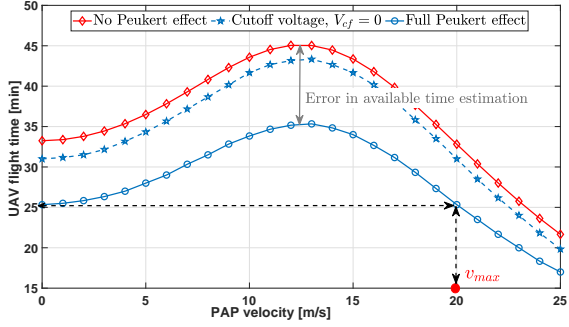}
\caption{Variation of available flight time with PAP velocity and cutoff voltage.}\label{fig:UAVvel}
\end{figure}

Fig.~\ref{fig:UAVvel} shows how the available flight time of a PAP is affected by its velocity, which is directly related to the power consumption. The flight time is maximized at 13 m/s because this velocity minimizes the UAV power consumption; then it decreases to the hovering level at 20 m/s. Above this velocity, the flight time is lower than in hovering conditions; the maximum velocity considered is 25 m/s because this allows to show this phenomenon while keeping the current under 10A per cell, the upper limit for this kind of batteries. The figure also shows the relevance of the Peukert effect by comparing the case where it is considered (blue curves) with one where it is neglected (red curve). In the latter, the flight time is calculated by simply dividing the battery capacity by the power consumption \cite{rui1}, \cite{babu3}. Since they result from a more conservative model, the blue curves are entirely below the red curve by a considerable margin, which is nearly constant when the cutoff voltage $V_{\text{cf}}$ is 0 (dashed blue curve). With $V_{\text{cf}}=3.2V$ (continuous blue curve), the gap with the red curve is variable because the secondary break condition in Algorithm \ref{algorithm2} ($V_{j}<V_{\text{cf}}$) is prevalent, causing a non-linear relation with the power consumption. The hovering time, assuming a UAV weight of 2 kg (excluding the battery), is estimated to be about 25 minutes, which is a sensible value for a commercial UAV\footnote[2]{\url{https://dl.djicdn.com/downloads/m100/M100_User_Manual_EN.pdf}}. 
\begin{algorithm}[]
\caption{Available Flight Time Estimation }
\label{algorithm2}
Initialize $P_{\text{uav}}$, $j=1$, $E_{\text{tot}}$, $E_{\text{max}}(I_\mathrm{b}^j)$, $V_\mathrm{b}^{j}$, $V_{\text{cf}}$, $I_\mathrm{b}^{j}$, $\Delta t$, $E_\mathrm{b}^{j}$\\
 \While{$E_{\text{tot}} < E_{\text{max}}(I_\mathrm{b}^j)$}
 {
    Calculate Slope $k$ with \eqref{slope};\\
    $V_\mathrm{b}^{j+1} = V_\mathrm{b}^j - k(I_\mathrm{b}^j)\cdot {E_\mathrm{b}^{j}}$;\\
    \If{$V_\mathrm{b}^{j+1} < V_{\text{cf}}$}{
    break;}
    $j=j+1$;\\
    $I_\mathrm{b}^{j}={P_{\text{uav}}}/{\left(V_\mathrm{b}^{j} n_{\mathrm{c}}\right)}$;\\
    $E_\mathrm{b}^{j} = \left(I_\mathrm{b}^{j}V_\mathrm{b}^{j}\right)\Delta t$;\\
    $E_{\text{tot}} = \sum_{t=1}^j E_\mathrm{b}^{t} $;\\
    $E_{\text{max}}(I_\mathrm{b}^{j}) = F_{\mathrm{E}}\left(I_\mathrm{b}^{j}\right)$;\\
 }
 \textbf{Output}:\,{$T(P_{\text{uav}}) = j\Delta t$.}
\end{algorithm}
\subsection{Trajectory Design Problem Formulation}\label{prblm}
The GEE of a PAP system can be increased by increasing the total number of bits transmitted and/or by reducing the PAP energy consumption with an efficient trajectory design. The flying velocity of the PAP affects: a) the PAP energy consumption; b) the discharge profile of its on-board battery
; c) the number of bits transmitted, which is a function of the time the PAP spends in the $m$-th path segment, $T_m$. The GEE trajectory optimization problem is formulated as,
\begin{IEEEeqnarray}{rCl}
\text{(P1)} & : & \underset{\{\mathbf{p}_{m}\},\{{T}_{m}\},\{{T}_{mn}\},\{{\mathbf{\Theta}}_{\mathrm{r}}^{m,i,n}\}}{\text{maximize}}\,\,\,\, \dfrac{\sum_{m=1}^{M}\sum_{n=1}^{N}T_{mn}D_{\mathrm{pg}}^{m,n}}{\sum_{m=1}^{M}{T}_{m}P_{\text{uav}}\left(v_{m}\right)}, \nonumber\label{p1}\\
\text{s.t.} & & \sum_{m=1}^{M}T_{mn}D_{\mathrm{pg}}^{m,n} \geq Q \quad \quad \forall n, \label{c1}\\
& & -\pi \leq \theta^{k}_{m,i,n} < \pi \quad \quad \quad\forall m, i, n, k, \label{c3}\\
& & \mathbf{p}_{{M+1}} = \mathbf{p}_{\mathrm{F}}; \mathbf{p}_{1} = \mathbf{p}_{\mathrm{I}},\label{c4}\\
& & T_{m} \geq 0; \quad T_{mn} \geq 0 \quad  \quad \forall m,n,\label{c5}\\
&&  b_{\mathrm{pr}}^{m,i},b_{\mathrm{rg}}^{i,n} \in \{0,1\}, \quad \quad \forall m, i, n,\label{c6}\\
& &\sum_{j=1}^{m}T_{j}P_{\text{uav}}(v_j)\leq E_{\text{max}}(P_{\text{uav}}(v_j)) \quad \forall m,\label{c9}\\
& &V_\mathrm{b}^m(P_{\text{uav}}(v_j))\geq V_{\text{cf}} \quad \forall m,\label{c10}\\
&& \eqref{path_descritization}, \eqref{tdma}.
\end{IEEEeqnarray}
The objective function of (P1) is the GEE of the system, the denominator of which is the total energy consumed by the aerial vehicle. Constraint \eqref{c1} demands that the PAP delivers $Q$ bits of data to the GNs by the end of the trajectory; the phase shift constraint associated with the IRS elements is represented by \eqref{c3}, whereas \eqref{c4} restrains the initial and final locations of the PAP; \eqref{c5} is the non-negative time constraint whereas \eqref{c9} and \eqref{c10} summarize the Peukert effect.

The solution to (P1) is not trivial, mainly due to the following reasons: a) the numerator of the objective function of (P1) has binary variables b) the denominator of the objective function and constraint \eqref{c1} are non-convex functions of the trajectory variables; c) the optimal design of phase shifts associated with the IRS elements to maximize the received SNR and the trajectory of the PAP are interlinked; d) the non-tractable form of the constraints \eqref{c9} and \eqref{c10}.

To tackle the above issues, we propose a two-phase algorithm: in the first phase, we determine a discretized GEE path that connects a set of \textit{locations-of-interest (LoIs)}, $\{\mathbf{p}_{m}\}\,\forall m\in \mathcal{M}$; in the second phase, the determined $\{\mathbf{p}_{m}\}$ are used to obtain the values of binary variables $\{b_{\mathrm{pr}}^{m,i}\}$, $\{b_{\mathrm{rg}}^{i,n}\}$. This allows us to accurately tune the phase shift values of the IRS reflecting elements so as to maximize the received SNR values at the GNs using an alternate optimization algorithm. \color{black} We then propose a novel multi-lap trajectory design method to efficiently allocate time to the GNs, while considering the Peukert constraints.  
\subsection{Phase 1: PAP Path Design and IRS beamforming}\label{Phase 1}
\begin{figure}{}
\centering
\includegraphics[width=0.85\columnwidth]{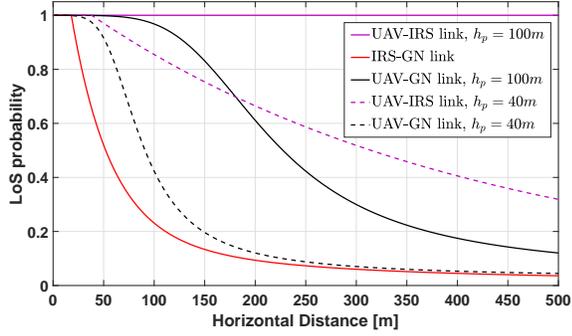}
\caption{\color{black}LoS probability variation of UAV-IRS, IRS-GN and UAV-GN links, with respect to projected horizontal distance.}\label{figplos}
\end{figure}
\begin{figure}[ht]
\centering
\begin{subfigure}[b]{0.49\linewidth}
\includegraphics[width=\textwidth]{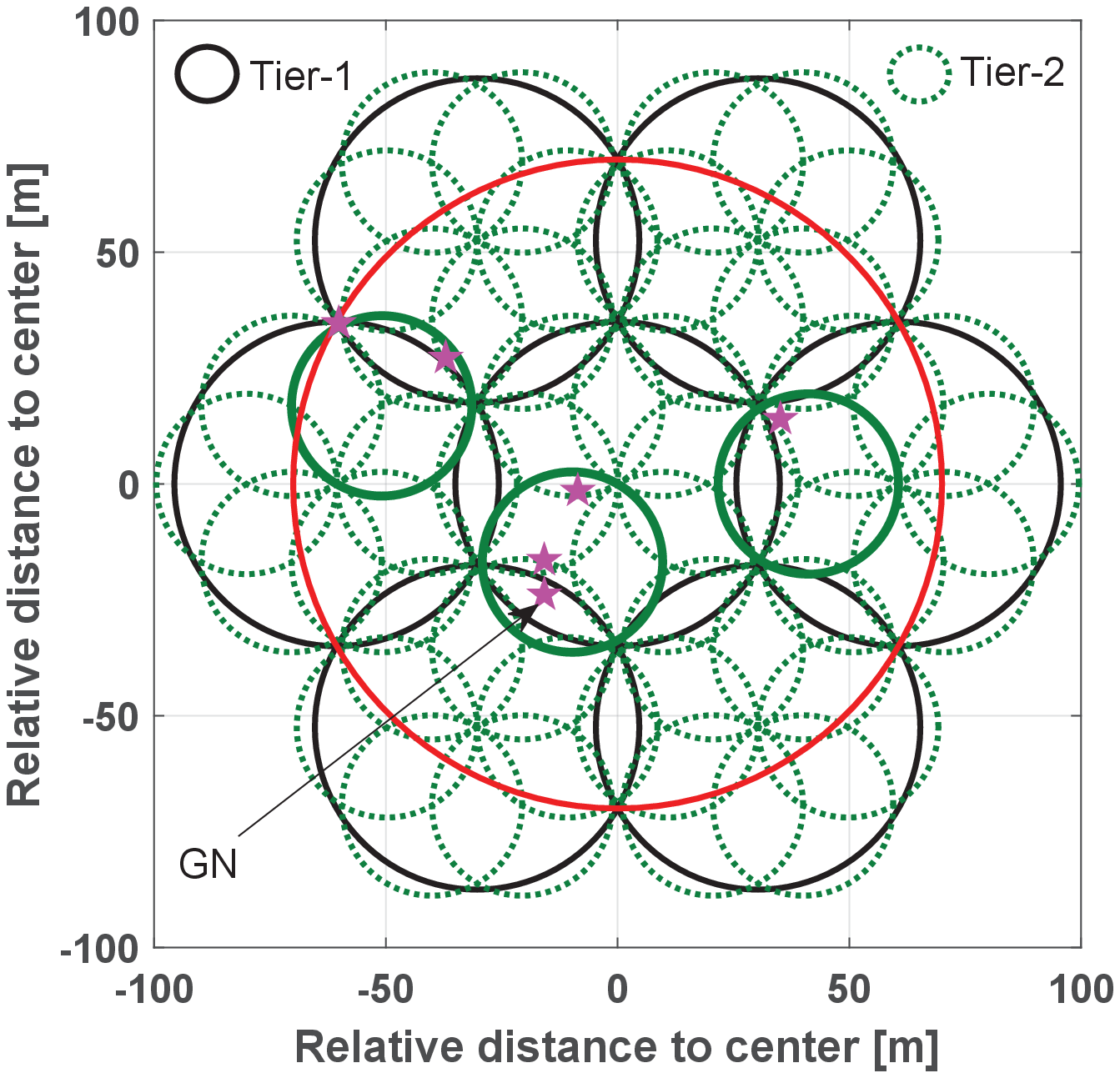}
\caption{ }\label{figloi}
\end{subfigure}
\hfill
\begin{subfigure}[b]{0.49\linewidth}
\includegraphics[width=\textwidth]{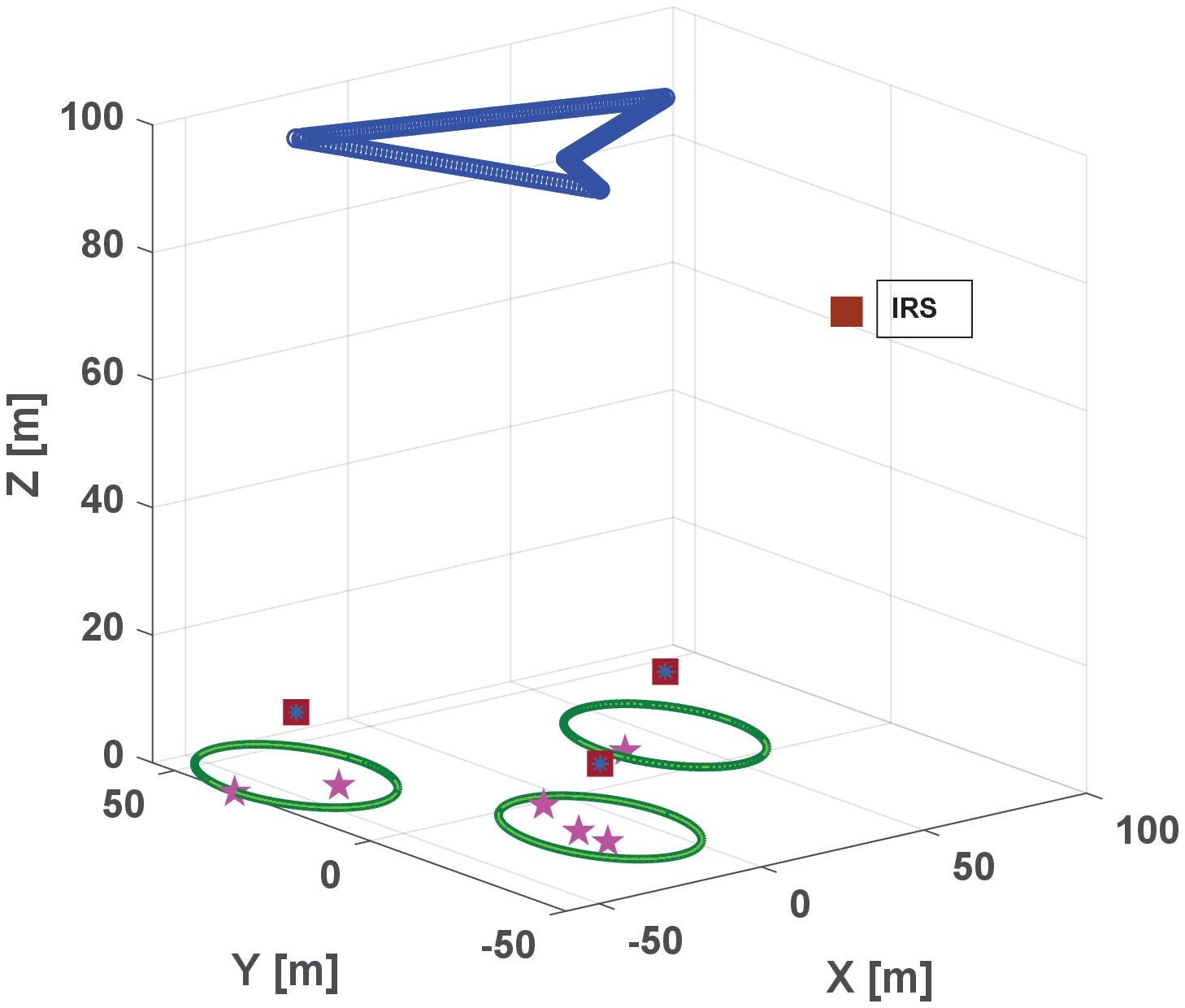}
\caption{ }\label{pappath}
\end{subfigure}
\caption{a)Set of LoS circles covering the users; b) PAP path through the LoIs.}
\end{figure}
\color{black}
\subsubsection{PAP Path Design and IRS positioning}
\begin{table}[]
\renewcommand*{\arraystretch}{1.6}
\caption{{\color{black}Packing Patterns \cite{circle}}.}
\centering
\begin{tabular}{|l|l|}
\hline
\color{black}{Number of circles, $u$} &\color{black} $R^{{u}}_{\text{max}}=\Lambda(u)\cdot R_{\mathrm{small}} $
\\\hline
\color{black}1 & \color{black} 1 $\cdot$ $R_{\mathrm{small}} $ \\\hline
 \color{black}2& \color{black}1 $\cdot$ $R_{\mathrm{small}} $ \\\hline
\color{black} 3& \color{black}2/$\sqrt{3}$ $\cdot$ $ R_{\mathrm{small}}$ \\ \hline
 \color{black}4&\color{black} $\sqrt{2}$ $\cdot$ $ R_{\mathrm{small}}$ \\\hline
 \color{black}5& \color{black}1.641 $\cdot$ $ R_{\mathrm{small}}$ \\\hline
 \color{black}6& \color{black}1.7988 $\cdot$ $R_{\mathrm{small}}$ \\\hline
 \color{black}7& \color{black}2 $\cdot$ $R_{\mathrm{small}}$ \\\hline
 \color{black}$u=8,9,10$& \color{black} $\left[1+2\text{cos}(2\pi/(u-1))\right]\cdot R_{\mathrm{small}}$ \\\hline
\end{tabular}\label{table1}
\end{table}
\color{black}The gain achieved by using IRSs to aid the communication between the PAP and a GN is expected to be significant when the PAP-IRS and IRS-GN are LoS links \cite{emil1}. Fig.~\ref{figplos} shows the variation of LoS probabilities between the PAP and a GN, the PAP and an IRS, and an IRS and a GN, obtained using \eqref{plospg}, \eqref{plospr}, and \eqref{plosrg}, respectively. For the considered urban scenario, a flying altitude of 100 m always guarantees a LoS link between the PAP and an IRS module \cite{3GPPlte}. Hence, we select the flying altitude $h_\mathrm{p}=100$ m. Additionally, a LoS link between an IRS and a GN can be guaranteed by placing the IRS module at a 2D distance of 20 m from the GN. Consequently, we cover the given geographical area of radius $R_{\mathrm{g}}$ by placing a set of small circles of radius 20 m using the proposed multi-tier packing algorithm.\\ \color{black}
\underline{Multi-tier Packing Algorithm:}
The multi-tier packing algorithm is an extension of the multi-level circle packing algorithm that considers the packing of 5 circles only in each level [Algorithm 2, \cite{babu3}]. If $R_{\mathrm{g}} > 20$, we need multiple smaller circles to cover the given region. Let $R^{{u}}_{\text{max}}$ be the maximum radius of the geographical region that $u$ small circles of radius $R_{\mathrm{small}}$ can cover. With the available packing patterns, in Table \ref{table1}, the maximum radius of the geographical region that can be covered using 10 circles is $2.53\cdot R_{\mathrm{small}}$. In practice, the considered region could be very large in dimension compared to the radius of the smaller circle ($R_{\mathrm{g}}>> R_{\mathrm{small}}$). Hence, our objective is to find the minimum number of smaller circles and the corresponding locations of their centers required to cover the given geographical area. The multi-tier packing concept is better explained in Fig.~\ref{figloi}: in the first tier of packing, 7 circles of radius $R_{\mathrm{g}}/\Lambda(7)$ are placed using 7-circle packing; in the second tier, each of these 7 circles is covered by 6 smaller circles of radius $R_{\mathrm{g}}/[\Lambda(7)\Lambda(6)]$, using 6-circle packing.\\
\underline{Proposition 1:}  The optimal circle packing pattern that requires the least number of circles of radius $R_{\text{small}}$ to cover a region of radius $R_t$ is determined as,
\begin{IEEEeqnarray}{rCl}
u^t_{\text{opt}} &=& \underset{u=1,2,..,10}{\text{argmin}} u^{\mu(u)},
\end{IEEEeqnarray}
where $\mu(u)={\dfrac{1}{\text{log}_{2}(\Lambda(u))}\text{log}_{2}\left(\dfrac{R_{t}}{R_{\mathrm{small}}}\right)}$, and $\Lambda(u)$ is obtained from Table II.\\
\underline{Proof:} The proof is a direct extension of the proof of Proposition 2 of \cite{babu3}.\\
\begin{algorithm}[]
 \label{algo_packing}
\caption{\color{black}Multi-tier packing algorithm.}
\color{black}
\textbf{Input}:\,$t=1$ $R_{t}=R_{\mathrm{g}}$, $R_{\mathrm{small}}$;\\
\While{$R_{t} \leq R_{\mathrm{small}}$}
{
Find the optimal packing pattern for the $t^{\text{th}}$ tier using Proposition 1;\\
Store the locations of the center of the circles to $\mathbf{l}_{t}$;\\
t=t+1;\\
Update the radius $R_t=R_{t-1}/\Lambda({u^{t-1}_{\mathrm{opt}}})$\\
}

\textbf{Output}:\,{$\mathcal{L}=\mathbf{l}_{t-1}$, the set of locations of the smaller circles to cover the given geographical area.}
\end{algorithm} 
\color{black}
Algorithm 2 gives the steps to follow to complete the multi-tier packing procedure. To find the path for the PAP, as shown in Fig.~\ref{figloi}, we first determine a set of locations using Algorithm 2 (center of green circles) with $R_{\mathrm{small}}=20$m: $\{\mathbf{l}_{t-1}\}\equiv\mathcal{L}$, that cover the given geographical area (red circle) entirely using the multi-tier packing method. Next, we consider the \textit{LoIs} as a subset of $\mathcal{L}$: $\mathcal{L}^{'}\subset \mathcal{L}$ for which each of the corresponding circles covers at least one GN (set of solid green circles):
\begin{IEEEeqnarray}{rCl}
\mathcal{L}^{'}& \equiv & 
\{\mathbf{l}_{{t-1}}\} \,\,\,\text{s.t}\,\, \|\mathbf{l}_{t-1}-\mathbf{g}_{n}\|\leq 20 \,\,\,\text{for at least one GN}.\IEEEeqnarraynumspace
\end{IEEEeqnarray}
Using $\mathcal{L}^{'}$, an energy-efficient path to cover the GNs in the geographical region is determined by finding the shortest path between points $\{\mathbf{l}_o\} \in \mathcal{L}^{'}$, starting from $\mathbf{p}_{\mathrm{I}}$ and ending at $\mathbf{p}_{\mathrm{F}}$ (constraint \eqref{c4}). The continuous path is then discretized into segments of length $\Delta$ (constraint \eqref{path_descritization}), giving a set of way points $\{\mathbf{p}_{m}\}$, as shown in Fig.~\ref{pappath}. For a set of way points, the optimum  $\mathbf{\Theta}_\mathrm{r}^{m,i,n} \ \forall \{m,i,n\}$ that maximizes the achievable received SNR at a GN can be found as shown below.
\subsubsection{\underline{IRS Beamforming Design}}
\color{black}

\color{black} We consider that the additional phase shift value introduced by an IRS element is limited to a discrete set of phase values due to practical hardware constraints. The possibility of using the $i^{\text{th}}$ IRS module to aid the communication between the PAP at the $m^{\text{th}}$ path segment and the $n^{\text{th}}$ GN is determined by the values of $b_{\mathrm{pr}}^{m,i}$ and $b_{\mathrm{rg}}^{i,n}$. If these are equal to one, the phase of each IRS element should be selected to maximize the GEE value. For a given $\{\mathbf{p}_{m}\},\{{T}_{m}\},\{{T}_{mn}\}$, from \eqref{snr} and \eqref{Dpg}, the optimal $\{{\mathbf{\Theta}}_{\mathrm{r}}^{m,i,n}\}$ that maximizes the GEE is the one that maximizes the SNR. Ideally, the phases must be adjusted to ensure constructive addition of the signals received through the direct and indirect paths from the PAP at a GN. However, due to the interdependence of the amplitude and phase responses of the IRS elements, as given in \eqref{amp-phase}, an additional phase shift introduced by an IRS element that guarantees a constructive addition of the received signals might produce a low amplitude response. Hence, an approximate solution can be determined by using the alternate optimization (AO) proposed in \cite{rui_amp}. The AO algorithm finds an approximate solution that maximizes the GEE by iteratively optimizing the phase shift of one of the K reflecting elements with those of the others being fixed at each time, and repeating this procedure for all K elements of an IRS module until the GEE value converges \cite{rui_amp}. This process must be repeated for all the IRS modules that satisfy $b_{\mathrm{pr}}^{m,i}=b_{\mathrm{rg}}^{i,n}=1$. Due to the considered discrete set of available phase values, the convergence of the algorithm is guaranteed. Using the optimal phase values obtained using AO, the value of $D_{\mathrm{pg}}^{m,n}$ $\forall m,n$ can be determined using \eqref{snr} in \eqref{Dpg}.
\subsection{Phase 2: Multi-Lap Trajectory Design}\label{phase2}
\color{black}In this phase, using the set of way points $\{\mathbf{p}_{m}\}$ and the determined $D_{\mathrm{pg}}^{m,n}$ values, we design a novel multi-lap trajectory that maximizes the GEE, while scheduling the GNs so that $Q$ bits of data are delivered to each of these by the end of the trajectory. Let us define a lap of the trajectory as the discretized path from $\mathbf{p}_{\mathrm{I}}$ to $\mathbf{p}_{\mathrm{F}}$ through all the LoIs.  The PAP takes $N_{\text{lap}}$ laps to deliver $Q$ bits of data to the GNs.  
For a given flying velocity $v_m$, the GNs can be scheduled by solving the following sub-problem of (P1), 
\begin{IEEEeqnarray}{rCl}
\text{(P1.1)} & : & \underset{N_{\text{lap}},\{{T}_{mn}\}}{\text{maximize}}\,\,\,\, \dfrac{Q}{N_{\text{lap}}\sum_{m=1}^{M}{T}_{m}P_{\text{uav}}\left(v_{m}\right)}, \nonumber\\
\text{s.t.} & & \sum_{m=1}^{M}T_{mn}D_{\mathrm{pg}}^{m,n} \geq Q/N_{\text{lap}} \quad \quad \forall n \in \mathcal{N},\label{P1.3c0}\\
&&  T_m = \dfrac{\Delta}{v_{{m}}}, \quad \quad \forall  m \in \mathcal{M}^{'}\label{P1.3c1}\\ 
&& N_{\text{lap}}+1 \leq \frac{T_{\text{max}}(v_m)}{M T_m}, \quad \quad \forall  \in \mathcal{M}^{'}\label{P1.3c2}\\
&&\eqref{tdma}, \eqref{c5} \label{P1.3c3}.
\end{IEEEeqnarray}
$T_{\text{max}}(v_m)$ is obtained in Algorithm 1 with the substitution $P_{\text{uav}}=P_{\text{uav}}(v_m)$, given by \eqref{puav}. (P1.1) is a convex optimization problem and can be solved using any available solvers, such as MATLAB's CVX. In practice, the velocity resolution of a UAV is determined by the on-board flight controller. Hence, we consider a finite set of velocity values that a UAV can take during its mission. (P1.1) is solved for different velocity values, and the one that maximizes the objective function is selected. Moreover, flying at a constant velocity avoids the additional power consumption linked to acceleration/deceleration, which is neglected by \eqref{puav}.  
 Algorithm \ref{E2P2} defines the overall energy-efficient path planning (E2P2) algorithm combining phases 1 and 2 of the proposed solution.
\begin{algorithm}[]
\caption{ $\text{E}2\text{P}2$ Algorithm}
\label{E2P2}
\color{black}Initialize the available phase set and the set of velocity values\\
Find the discretized path between the LoIs, as explained in Section~\ref{Phase 1}: $\{\mathbf{p}_{m}\}$;\\
Find $b_{\mathrm{pr}}^{m,i}, b_{\mathrm{rg}}^{i,n}$ $\forall i, n$ using $\{\mathbf{p}_{m}\}$ and $\{\mathbf{g}_{n}\}$;\\
Find the optimal amplitude-phase shift matrix of the IRSs ($ \{\mathbf{\Theta}_{\mathrm{r}}^{m,i,n}\} $) using the AO proposed in \cite{rui_amp};\\
Find $D_{\mathrm{pg}}^{m,n}$ $\forall m, n$ using \eqref{Dpg};\\
 \For{each $ v \in \mathcal{V}$}
 {
 $v_{m}=v, T_m=\frac{\Delta}{v_m}\quad \forall m$, calculate $P_{\text{uav}}(v_{m})$ and estimate $T_{\text{max}}(v_m)$ using Algorithm 1;\\
  Determine the optimal solution of (P1.1): $\{{T}_{mn}\}$, $N_{\text{lap}}$;\\
 \If{GEE does not improve}
 {
break;
 }
 }
 \textbf{Output}:\,{Optimal PAP trajectory variables: $ \{\mathbf{p}_m\}, \{{T}_{m}\},\{{T}_{mn}\}$,$\{\mathbf{\Theta}_{\mathrm{r}}^{m,i,n}\}$, $N_{\text{lap}}$.}
\end{algorithm}
\color{black}
\section{Numerical Analysis and Discussion}\label{result}
\begin{table}[]
\caption{Simulation Parameters}
\begin{tabular}{lll}
\hline
Label & Definition & Value \\ \hline
\hline
$B_c$ & Channel Bandwidth for each GN & 20 \color{black}MHz\color{black} \\
$\sigma^2$ & Noise Power & -101 dBm \\
$h_p$ & PAP's flying altitude & 100 m \\
$\beta$ & antenna beamwidth & $45^{0}$\\
$v_\text{max}$ & Maximum achievable PAP speed & 20 m/s \\
$\Delta$ & Path Discretization Interval & 1 m \\
$P$ & Transmission Power & 23 dBm \\
$(a,b)$ & LoS probability constants for Suburban topology & (4.88,0.43) \\
$\eta_1$ & additional mean pathloss for LoS group & 0.2 dB \\
$\eta_2$& additional mean pathloss for for NLoS group & 24 dB \\ \hline
\end{tabular}
\label{simulation}
\end{table}
In this section we provide the main findings through numerical evaluation. The parameters used for the simulation are given in Table \ref{simulation}.
\subsection{Battery Design}
\begin{figure}{}
\centering
\includegraphics[width=0.85\columnwidth]{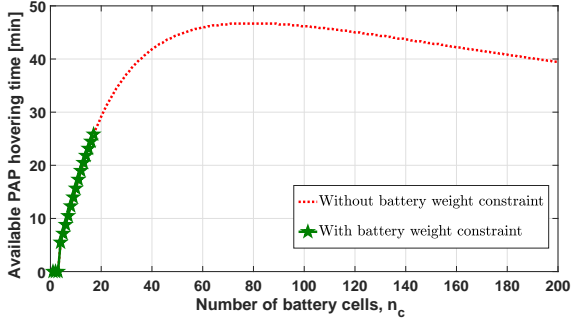}
\caption{Hovering time as a function of battery size, with and without weight constraint.}\label{fig:battdes}
\end{figure}

The model and methodology described in Section~\ref{bat_design} were used to study several aspects related to the battery design, such as the effect of flight velocity, the battery size and the relevance of the Peukert effect. As explained before, Fig.~\ref{fig:UAVvel} shows the relevance of the Peukert effect by comparing the case where it is considered with one where it is neglected. Fig.~\ref{fig:battdes} shows how the available hovering time varies with different numbers of battery cells. Under the hypothesis of unconstrained battery size, represented by the dotted line, it is possible to observe the trade-off between the extra capacity and the extra weight provided by each extra cell; the weight of a battery cell is considered as 50 g. The positive effect of a larger battery tends to provide decreasing marginal returns as the battery size grows, until the trend is reversed and a heavier battery starts having a negative impact on the hovering time. In reality, it is not possible to increase the UAV weight indefinitely, because there is a maximum take-off weight the motors can withstand, which is set at 3.6kg in this case (inclusive of UAV and battery). When such constraint is considered, the range of possibilities is restricted to the continuous curve in Fig.~\ref{fig:battdes}, and the hovering time maximization is achieved at its upper boundary, corresponding to 17 battery cells and an hovering time of about 25 minutes, as shown in Fig.~\ref{fig:UAVvel}.
Under the current assumptions, the battery configuration is irrelevant, as it is evident from step 8 of Algorithm~\ref{algorithm2}. Nonetheless, 17 is a prime number, so the only way to achieve it is to put all cells either in series or in parallel, which is an unrealistic solution. Therefore, a battery size of 16 cells is a more sensible option.
\subsection{IRS design}
\begin{figure}{}
\centering
\includegraphics[width=0.85\columnwidth]{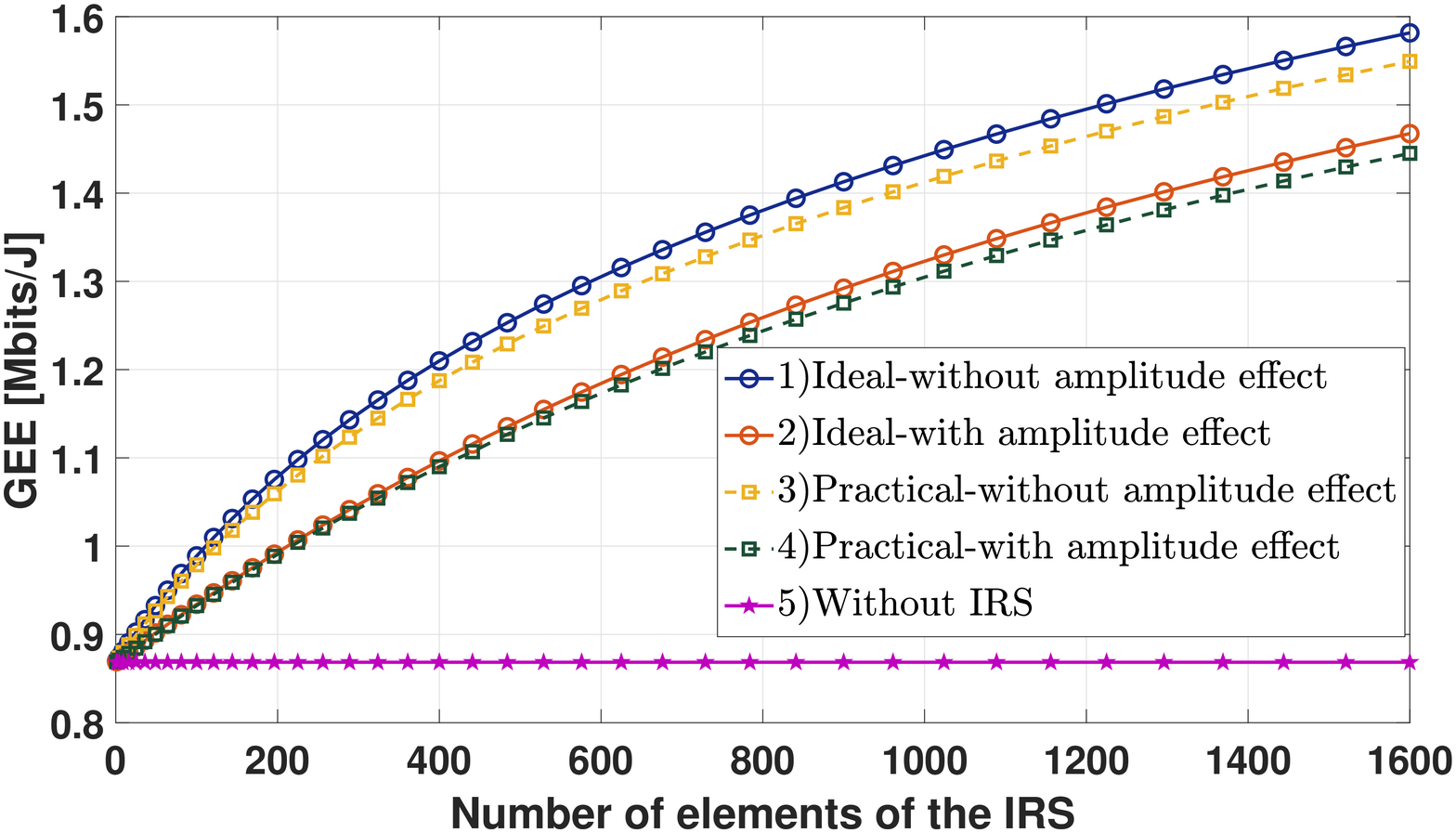}
\caption{\color{black}GEE vs Number of elements of the IRS.}\label{single_irs}
\end{figure}
\color{black}
Fig.~\ref{single_irs} represents the variation of GEE as a function of the number of elements of a single-IRS-single-user system. The user is placed in the NLoS regime of the PAP position. The figure shows five different cases:
\begin{itemize}
    \item Case 1: GEE achieved if the IRS phase values are optimized neglecting the amplitude and phase dependency;
    \item Case 2: The actual received GEE value with the phase optimized according to Case 1;
    \item Case 3: Each element of the IRS can take any of the available four options: $\{0,\pi/2,-\pi/2,\pi\}$, and the phases are optimized without considering the amplitude effect using the alternate optimization algorithm;
    \item Case 4: Each element of the IRS has only 4 phase shift options $\{0,\pi/2,-\pi/2,\pi\}$ and the phases are optimized considering the amplitude effect using the alternate optimization algorithm;
    \item Case 5: GEE of the system without IRS.
\end{itemize}
As expected, the GEE value improves with the number of elements of the IRSs due to an increase in the number of PAP-IRS and IRS-GN paths. An IRS of area 0.25 $m^2$ improves the GEE by 63\% when the user is in the NLoS regime of the PAP. The figure also shows the effect of considering the interdependence of amplitude and phase responses of the IRS element. Optimizing the IRS phase shift values without considering the amplitude effect could lead to an over-estimation error of 10\%. This might result in a trajectory that does not allow the PAP to reach the goal of transmitting $Q$ bits of data to all users, since these are scheduled based on the overestimated spectral efficiency values. Additionally, a discrete set of 4 phase values could achieve a performance comparable to an ideal scenario that allows the IRS elements' phases to be tuned to any value in the interval $\left[-\pi,\pi\right)$.
\subsection{PAP trajectory design}
\begin{figure}{}
\centering
\includegraphics[width=0.85\columnwidth]{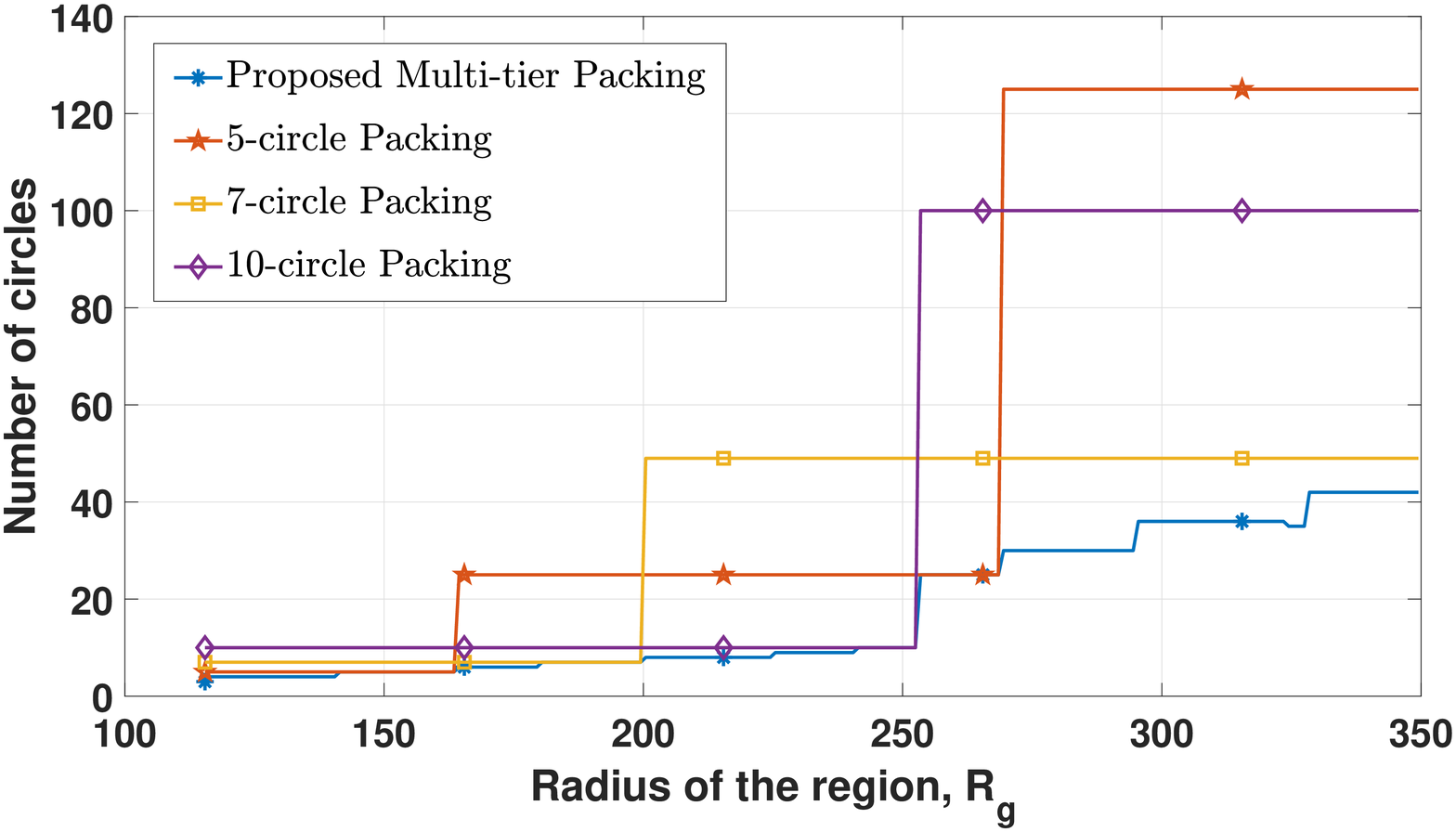}
\caption{\color{black}Comparison of different packing algorithms.}\label{fig:pack_comparison}
\end{figure}
\color{black}Fig.~\ref{fig:pack_comparison} compares the total number of circles required to cover a given geographical area according to the proposed multi-tier packing algorithm, with those using 5,7,10 circles in each level of the multi-level circle packing algorithm, as proposed in \cite{babu3}. As the figure shows, the proposed method always guarantees the least number of circles, even by a conspicuous margin when the radius of the region to be covered is three or more times the radius of the smaller circle. The radius of the smaller circle, $R_{\text{small}}$, is taken as 100 m. 
\color{black}Now, we compare the proposed multi-lap trajectory design with two base scenarios:
\begin{itemize}
    \item Baseline 1: GEE obtained if the PAP follows a fly-hover-communicate protocol to serve the GNs \cite{babu3};
    \item Baseline 2: GEE obtained if the mission is to be completed in a single lap \cite{rui1}. \end{itemize}
\subsubsection{Single lap trajectory design}
In this baseline, the PAP is expected to complete the mission in a single lap. The corresponding optimization problem can be written as, 
\begin{IEEEeqnarray}{rCl}
\text{(P1.2)} & : & \underset{\{{T}_{m}\},\{{T}_{mn}\}}{\text{maximize}}\,\,\,\, \dfrac{\sum_{m=1}^{M}\sum_{n=1}^{N}T_{mn}D_{\mathrm{pg}}^{m,n}}{\sum_{m=1}^{M}{T}_{m}P_{\text{uav}}\left(v_{m}\right)}, \nonumber\\
\text{s.t.} & & \dfrac{\Delta}{T_m} \leq v_{\text{max}}, \quad \quad \forall  m \in \mathcal{M}^{'},\label{P1.1c1} 
\end{IEEEeqnarray}
\begin{IEEEeqnarray}{rCl}
&& \sum_{m=1}^{M} T_{m} \leq T_{\text{max}}, \label{P1.1c2}\\
&&\eqref{tdma}, \eqref{c1}, \eqref{c5} \label{P1.1c3}.
\end{IEEEeqnarray}
Using \eqref{P1.1c1} and \eqref{P1.1c2}, the Peukert constraints are equivalently represented by limiting the PAP flying velocity to $v_{\text{max}}$, such that $P_{\text{uav}}(v_{\text{max}})\leq P_{\text{uav}}(0) $. Furthermore, the total trajectory time is constrained to be less than $T_{\text{max}}$; $T_{\text{max}}$ is the maximum available time if the PAP is hovering, determined using Algorithm \ref{algorithm2}, as marked in Fig.~\ref{fig:UAVvel}. Consequently, (P1.2) takes the form of a fractional programming problem and can be solved using the Generalized Dinkelbach's Algorithm if the numerator is a concave function and the denominator is a convex function of the optimization variables \cite{babu1}. Also, all the constraints have to be convex in nature.

The $P_{\text{uav}}(v_m)$ given by \eqref{puav} makes the denominator term of the objective function of (P1.2) a non-convex function of the optimization variable $T_m$. We use the sequential convex programming (SCP) approach to tackle the non-convex objective function. The fundamental idea of the SCP technique is to solve iteratively a sequence of convex approximated problems of the original non-convex problem, so that the feasible solution points converge to the KKT point of the original non-convex problem \cite{babu1}.

Let,
\begin{IEEEeqnarray}{rCl}
E(T_m)&=& T_m P_{\text{uav}}\left(\dfrac{\Delta}{T_{m}}\right),\\
&=& C_1\left(T_m+\dfrac{3\Delta^2}{{T_m}v_{\text{tip}}^2}\right)+C_2 \dfrac{\Delta^3}{{T_m}^2}\nonumber\\
&+& C_3\left(\sqrt{C_4 {T_m}^4+\dfrac{\Delta^4}{4}}-\dfrac{\Delta^2}{2}\right)^{1/2},\label{D}
\end{IEEEeqnarray}
with $\Delta=\|\mathbf{p}_{m+1}-\mathbf{p}_{m}\|$ $\forall m\in \mathcal{M}^{'}$. By introducing the slack variable $z_m$ such that,
\begin{IEEEeqnarray}{rCl}
&&{z^2_m}=\left(\sqrt{C_4 {T_m}^4+\dfrac{\Delta_m^4}{4}}-\dfrac{\Delta^2}{2}\right),\label{zmj}\label{slack2}
\end{IEEEeqnarray}
where $C_1=N_R P_b$, $C_2=\dfrac{1}{2}C_{D}A_{\text{f}}\rho(h_\mathrm{p})$, $C_3=W$, and $C_4={W^2}/\{4 N_{\mathrm{R}}^2 \rho^2 (h_\mathrm{a})A_{\mathrm{r}}^2\}$. By substituting \eqref{slack2} in \eqref{D}, \eqref{D} can be written in the convex form as,
\begin{IEEEeqnarray}{rCl} \label{stm}
E(T_m) & = & C_1\left(T_m+\dfrac{3\Delta^2}{{T_m}v_{\text{tip}}^2}\right)+C_2 \dfrac{\Delta^3}{{T_m}^2}+ C_3z_{m}, \label{etm}
\end{IEEEeqnarray}
such that
\begin{IEEEeqnarray}{rCl}
&& \dfrac{{T_{m}}^4}{{z^2_m}} \leq \dfrac{{z_m^{l}}^2+2z_m^{l}\left(z_m-z_m^{l}\right)+\Delta^2}{C_4},\label{con2}
\end{IEEEeqnarray}
where \eqref{con2} is the first order Taylor expansion of \eqref{slack2} around the point $z_{m}^{l}$.
Hence, the optimization problem (P1.2) can be rewritten as,
\begin{IEEEeqnarray}{rCl}
(\text{P1.3})& & \underset{\{{T}_{m}\},\{{T}_{mn}\}}{\text{maximize}}\,\,\,\, \dfrac{\sum_{m=1}^{M}\sum_{n=1}^{N}T_{mn}D_{pg}^{m,n}}{\sum_{m=1}^{M}E(T_m)} \nonumber\\
\text{s.t} &  & \,\,\,\,\eqref{P1.1c1}-\eqref{P1.1c3}, \eqref{con2}.
\end{IEEEeqnarray} 
The above problem is convex and can be solved using Dinkelbach's algorithm. To find the optimal trajectory, steps 6-10 of Algorithm 3 are replaced with the Generalized Dinkelbach's algorithm to solve (P1.3).
\begin{figure}[ht]
\centering
\includegraphics[width=0.85\columnwidth]{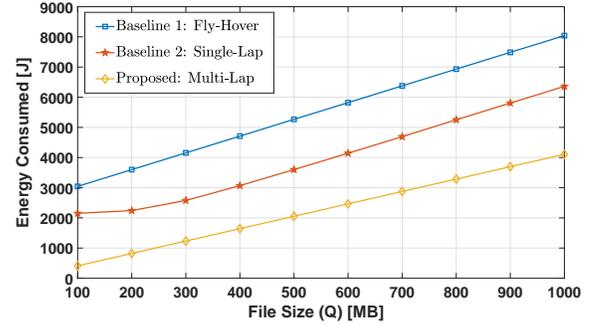}
\caption{\color{black}Total energy consumed by the PAP as a function of the file size to be delivered to each GN.}\label{figtraj}
\end{figure}

Fig.~\ref{figtraj} shows the amount of energy saved by having the the PAP follow the proposed multi-lap trajectory, compared to single-lap and fly-hover communicate policies. The solution is obtained after executing the E2P2 algorithm considering a set of 6 GNs uniformly distributed over a circular area of radius 60 m. Out of the LoS coverage circles obtained using the multi-tier packing algorithm (Algorithm 2), 3 LoIs are selected that cover the GNs. The shortest path between them, starting and ending at (0,0,100), is determined. In the fly-hover-communicate policy, the PAP hovers at LoIs to serve the GNs with a file of a given size. The PAP is assumed to fly with the maximum velocity between the LoIs. In the single-lap baseline scenario, the velocity with which the PAP covers a path segment and the scheduling of the GNs are calculated by solving (P1.3) with the Generalized Dinkelbach's algorithm \cite{babu1}. The initial feasible value $z_m^l$ to solve (P 1.3) is taken as the solution of the fly-hover-communicate policy. The amount of energy consumed while following a multi-lap policy is always lower than in the two other baseline scenarios, and the gain scales with the size of the file to be delivered to the GNs. This is because the single-lap policy involves hovering at LoIs for longer to complete the data transmission. In contrast, the multi-lap policy allows the PAP to deliver a file in batches over multiple laps. Since the power consumed during hovering is greater than flying horizontally, the multi-lap policy requires less energy than the single-lap counterpart. The fly-hover-communicate policy is the most energy-hungry, but it has lower complexity than the other two policies.
\color{black}
\section{Conclusion}
In this work, we proposed an algorithm to design a GEE trajectory for a multi-IRS assisted PAP deployed to deliver a given amount of data to a set of GNs, while taking into account the non-linear discharge behavior of the PAP battery. Furthermore, an algorithm to estimate the available flight time of a PAP for different flying velocities has been provided. The proposed two-phase GEE PAP trajectory design solution allows to consider the interdependence of the phase and amplitude responses of the IRS modules on the received SNR. From the numerical evaluation, it is observed that: adding more battery cells to a PAP battery unit does not always increase the available flight time, since it also increases the weight of the PAP; \color{black}neglecting the amplitude-phase dependency of IRS elements leads to an overestimation of the GEE of the system. \color{black} Finally, a fly-communicating PAP system always has a higher GEE compared to the fly-hover-communicate counterpart. The presence of IRS modules enhances the GEE by providing extra separate paths for the signal to reach a GN from the PAP. 

The trajectory design for a multi-UAV system, with some UAVs carrying IRS modules on-board and the remaining ones configured as PAPs deployed to serve a set of moving GNs, is left as future work. \color{black} The same is true for models taking into account the effect of variable temperature on the energy availability of the battery.\color{black}

\bibliographystyle{IEEEtran}
\bibliography{./main.bbl}
\end{document}